\documentclass[final]{siamart1116}

\usepackage{lipsum}
\usepackage{amsfonts}
\usepackage{graphicx}
\usepackage{epstopdf}
\usepackage{algorithmic}
\ifpdf
  \DeclareGraphicsExtensions{.eps,.pdf,.png,.jpg}
\else
  \DeclareGraphicsExtensions{.eps}
\fi

\numberwithin{theorem}{section}

\newcommand{\TheTitle}{Singularity of a combustion wave profile: a clue to the multi-component theory for liquid-gas filtration} 
\newcommand{\TheAuthors}{M. A. Endo Kokubun and A. A. Mailybaev}

\headers{Multi-component theory for liquid-gas filtration combustion}{\TheAuthors}

\title{{\TheTitle}\thanks{Submitted to the editors DATE.
\funding{This work was supported by the
CNPq (grant 302351/2015-9) and the Program FAPERJ Pensa Rio (grant E-26/210.874/2014).}}}

\author{
  Max Akira Endo Kokubun\thanks{Department of Chemistry, University of Bergen, Norway and Instituto Nacional de Matem\'{a}tica Pura e Aplicacada - IMPA, Rio de Janeiro, Brazil
    (\email{max.kokubun@uib.no}).}
  \and
Alexei Mailybaev\thanks{Instituto Nacional de Matem\'{a}tica Pura e Aplicada - IMPA, Rio de Janeiro, Brazil (\email{alexei@impa.br}).}
}

\usepackage{amsopn}

\ifpdf
\hypersetup{
  pdftitle={\TheTitle},
  pdfauthor={\TheAuthors}
}
\fi

\begin{document}

\maketitle

\begin{abstract}
We study a nonlinear wave for a system of balance laws in one space dimension, which describes combustion for two-phase (gas and liquid) flow
in porous medium.
The problem is formulated for a general $N$-component liquid for modeling the strong multi-component effects reported recently for an application
to light oil recovery by air injection.
Despite the immense complexity of the model, the problem allows analytic solution.
The clue to this solution is a special form of a folding singularity, which occurs at an internal point of the wave profile.
Analysis of this singularity provides a missing determining relation for wave parameters.
This result is not only interesting for the application under consideration, but also motivates a deeper mathematical study of such singularities
for general systems of balance laws.
\end{abstract}



\section{Introduction}

The use of air injection as a method of enhanced oil recovery has been explored for a long time \cite{sarathi1999situ,CastanierAUG2003}.
In this method, part of oil burns with the injected air, increasing the well temperature and lowering the oil viscosity, thus enhancing
its mobility.
Traditionally, air injection has been used to recover heavy oils, i.e., oils with a very high viscosity.
In this case, chemical reactions crack the oil into a non-volatile part (coke) and volatile components,
which are expelled from the high temperature region~\cite{farouq2003}.
The exothermic reaction between the coke and the oxygen leads to formation of a high-temperature oxidation (HTO) wave that propagates in the porous rock.
The method of air injection has also been reported to increase recovery rates of light oils \cite{falade1989,ren2002air,gargar2015jpse}.
In this case, thermal expansion and gas drive promoted by the oxidation reaction are responsible for enhancing the recovery of oil.
The reaction that takes place between light oil and injected oxygen occurs at lower temperatures, bounded by the boiling point; it is termed as low-temperature oxidation (LTO).
Multicomponent nature of oil is essential for the structure of the resulting combustion wave, as it was shown recently with a simplified two-component model~\cite{gargar2014jpm,kokubun2016}.
The aim of the present paper is to develop a mathematical theory for the liquid-gas filtration combustion in the full multicomponent formulation.

As opposed to the theory of HTO, which usually considers a single-phase gas flow \cite{schult1995propagation,schult1996forced,mailybaevEtAl2010b,chapiro2012asymptotic},
our problem is based on a two-phase flow model~\cite{mailybaev2011resonance,mailybaev2013recovery,kokubun2016}
because the liquid fuel is also mobile.
Furthermore, the mathematical formulation must take into account not only
oxidation, but also vaporization and condensation of the volatile fuel.
Even though a practical problem is intrinsically multidimensional, the use of one-dimensional models allows for the
analytical study of fundamental characteristics~\cite{BruiningEtAl2008,wahle2003effects}.
It was shown with a simplified (single pseudo-component) model that the wave parameters (speed and temperature) must be extracted from the analysis
of internal wave profile \cite{mailybaev2011resonance}, 
following from the conditions at a fold singularity, a so-called resonance point.
Physically, at a resonance point the combustion wave speed coincides with the saturation (Buckley--Leverett) characteristic speed.

In this paper we study traveling wave solutions for a system of balance laws in one space dimension, modeling two-phase reactive flow with an $N$-component liquid fuel.
The problem is reduced to finding a heteroclinic orbit, connecting 
two limiting constant states, of a vector field on a folded ($2N+1$)-dimensional manifold in configuration space. Since the limiting states are on different sides of the fold and a generic orbit cannot pass the fold, an extra singularity condition follows. 
Such a property provides a missing determining condition for the combustion wave parameters, reducing the solution to a system of explicit algebraic equations. For a specific problem, this system can be easily solved numerically, as we demonstrate with an example of 3-component oil model.

Note that traveling wave solutions with an internal singularity at the resonance point are known also as the pathological~\cite{fickett2011detonation} or canard~\cite{harterich2003viscous}  case. Their appearance in natural phenomena, though rather rare, ranges from astrophysics~\cite{sharpe1999structure} to combustion and thermodynamics~\cite{weiss1995continuous}. A mathematical theory of this singularity is developed for the simplest case of one balance law~\cite{harterich2003viscous}, but a completely understanding and classification are missing in the case of systems of balance laws. Note that the traveling wave under consideration connects distant zeros of source terms in balance laws, which distinguishes this case from hyperbolic conservation laws with relaxation~\cite{liu1987hyperbolic,zumbrun2000existence}.
The proposed methodology is closely related to the theory of fast-slow systems (geometric singular perturbation theory)~\cite{kuehn2015multiple} and its applications to the study of traveling wave solutions~\cite{wechselberger2010folds}.
We hope that our results will motivate further studies of this phenomenon both in the area of pure mathematics and applications.

The paper is organized as follows. In \cref{secModel} we present the model formulation and
in \cref{sec:CC} we identify the wave structure that arises for large times.
In \cref{sec:Conl} we obtain the unknown limiting states, while \cref{sec:wprof} is devoted to obtaining the internal wave profiles and determining the overall structure and fold singularity.
In \cref{sec.ONE} we describe the hypersurface for the case of a single-component oil, obtaining the missing condition at the singularity, and in \cref{sec.Ncomp} we generalize the results for the $N$-component case.
Finally, we provide a numerical example in \cref{sec:numer} and present the conclusions in \cref{sec.conc}

\section{Multi-component model for two-phase reactive flow in porous medium}
\label{secModel}

We consider a system of balance laws that describes the two-phase (liquid and gas) flow in porous medium, with reactions and phase transitions.
The liquid has an arbitrary number of components, $N$, representing a detailed model for complex fluids.
In this paper, we refer to the recovery of light oil by air injection as a major application,
where intense multi-component effects were reported~\cite{gargar2014jpm,kokubun2016}.

\subsection{Equations and variables}

We consider a one-dimensional problem with spatial coordinate $x$ and time $t$. As dependent variables, we consider a temperature $T$, a total Darcy velocity $u$, a saturation of liquid (oleic) phase, $0 \le s_o \le 1$ (with a saturation of the gaseous phase $s_g = 1-s_o$), and fractions of all components in each phase.
The oleic phase contains $N$ components, and we consider volume fractions for each component $X_1,\ldots,X_{N-1}$, with the last component described by $X_N = 1-X_1-\cdots-X_{N-1}$.
In the gaseous phase, we distinguish $N$ oil components with molar fractions $Y_1,\ldots,Y_N$ and the fraction of oxygen denoted by $Y_k$; the remaining gaseous components with fraction $Y_r = 1 - Y_k - \sum_{i=1}^N Y_{i}$ consist of reaction products and inert components of the injected gas. Note that volume fractions are the same as molar fractions in the ideal gas. We will adopt the notations $\mathbf{X} = (X_1,\ldots,X_{N-1})$ and $\mathbf{Y} = (Y_1,\ldots,Y_{N})$ for oil components. In summary, there are $2N+3$ dependent variables in the problem: $T$, $u$, $s_o$, $\mathbf{X}$, $\mathbf{Y}$ and $Y_k$.

Balance laws will be constructed for each oil and gas component and heat, with two types of source terms: phase transitions (vaporization or condensation) and reactions. We use the following simplified reaction model:
\begin{equation}
\nu_{o_i}\,(\textrm{liquid component } i) + \textrm{O}_2 \rightarrow \nu_{g_i}\,(\textrm{gaseous products}),
\quad i=1,2,\ldots,N,
\label{eqM.01}
\end{equation}
which means that one mole of oxygen reacts with $\nu_{o_i}$ moles of $i$th oil component in oleic phase generating $\nu_{g_i}$ moles of gaseous products such as H$_2$O, CO$_2$, etc.
Gaseous phase reactions are disregarded, which is typical for petroleum applications due to annihilation of free radicals at pore walls~\cite{levenspiel1999chemical}.

Balance laws are obtained as a direct generalization to the case of $N$ components of the two-component model proposed in~\cite{gargar2014jpm,kokubun2016}. The full system of $2N+3$ equations is
\begin{align}
\frac{\partial}{\partial t}
\varphi {X}_i \rho_i  s_o
+
\frac{\partial}{\partial x}\rho_i u_{o_i}
& =
-\nu_{o_i} W_{r_i} - W_{v_i}, \ \ \ i=1,\ldots,N;
\label{eqM.xi}
\\[5pt]
\frac{\partial}{\partial t}
\varphi Y_i \rho_g s_g
+
\frac{\partial}{\partial x}\rho_g u_{g_i}
& = W_{v_i}, \ \ \ i=1,\ldots,N;
\label{eqM.yi}
\\[5pt]
\frac{\partial}{\partial t}
\varphi Y_k\rho_g s_g
+
\frac{\partial}{\partial x}\rho_g u_{g_k}
& =
- \sum_{i=1}^N W_{r_i};
\label{eqM.yk}
\\[5pt]
\frac{\partial}{\partial t}
\varphi Y_r\rho_g s_g
+
\frac{\partial}{\partial x}\rho_g
u_{g_r}
& =
\sum_{i=1}^N \nu_{g_i} W_{r_i},
\label{eqM.yr}
\end{align}
\begin{equation}
\begin{array}{l}
\displaystyle
\frac{\partial }{\partial t}
\left(
(1-\varphi)C_m + \varphi C_o s_o + \varphi c_g\rho_g s_g
\right)
(T-T_{ini})
+
\frac{\partial}{\partial x}
\left(
C_o u_o + c_g\rho_g u_g
\right)
(T-T_{ini})
\\[10pt]
\displaystyle
\qquad =
\lambda\frac{\partial^2 T}{\partial x^2}
+
\sum_{i=1}^N
Q_{r_i} W_{r_i}
-
\sum_{i=1}^N
Q_{v_i} W_{v_i}.
\end{array}
\label{eqM.t}
\end{equation}
Here, Eq.~\cref{eqM.xi} is the balance law for the $i$th component in the oleic phase, where $\rho_i$ is the corresponding oleic molar density, $u_{o_i}$ is the Darcy velocity, and $\varphi$ is the rock porosity. The source terms $W_{r_i}(T,\mathbf{X},Y_k,s_o)$ and $W_{v_i}(T,\mathbf{X},\mathbf{Y},s_o)$ are the reaction and vaporization rates. Equation~\cref{eqM.yi} describes the balance law for the same component in the gas phase, where $u_{g_i}$ is the corresponding Darcy velocity and the reaction is neglected as we mentioned above. Equations~\cref{eqM.yk} and \cref{eqM.yr} correspond, respectively, to the oxygen component and to the remaining inert gas in the gas phase. Finally, Eq.~\cref{eqM.t} governs the heat balance, where $T_{ini}$ the initial reservoir temperature, $\lambda$ is the effective thermal conductivity, $C_m$, $C_o$ and $c_g$ are the heat capacities of the rock matrix, oil and gas, $Q_{r_i}$ and $Q_{v_i}$ are the reaction enthalpy and vaporization heat for the respective component. In these equations, the Darcy velocities, gas density and reaction/vaporization rates are given functions described below; the other parameters are assumed to be constant.

Several simplifications were used in our model: we neglected
water originally present or condensed from the reaction products since its effect is not very strong~\cite{gargar2014fuel}.
Also, we neglected heat losses, which are usually small in field applications.
The mass and capillary diffusion will not be considered.
For more details on physical assumptions and justifications, see~\cite{kokubun2016}.

\subsection{Darcy velocities and source terms}
\label{sec2.2}

\begin{figure}
 \begin{center}
  \begin{tabular}{cc}
  \includegraphics[width=0.4\linewidth]{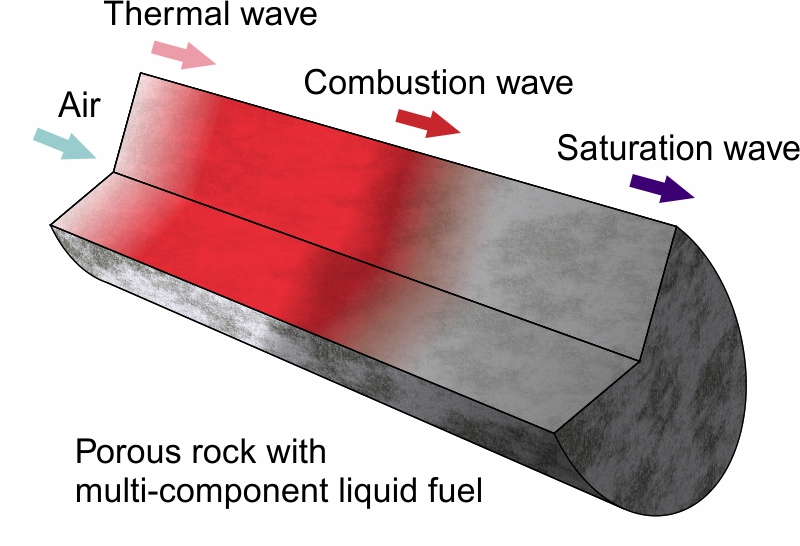} &
  \hspace{5mm}
  \includegraphics[width=0.25\linewidth]{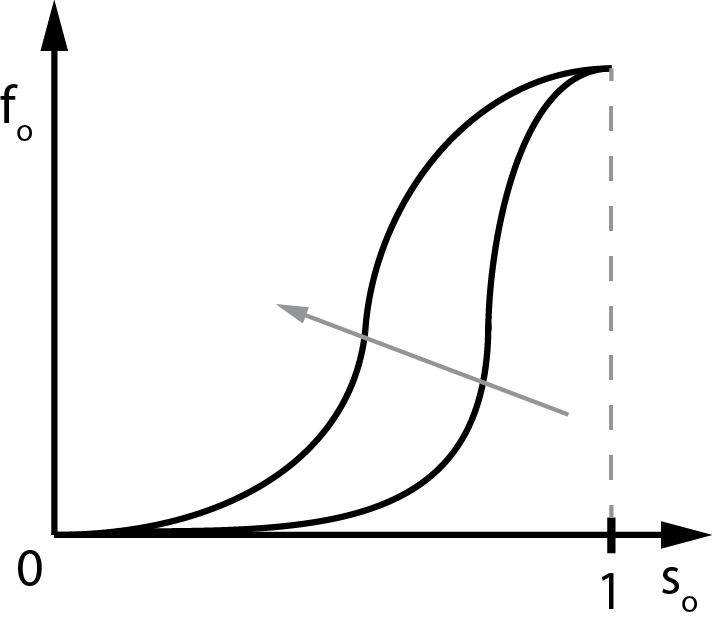}
  \end{tabular}
  \end{center}
  \caption{Wave sequence solution with the thermal, combustion, and saturation waves, where the red color indicates a hot region.
  The right panel shows typical shapes of the fractional flow function $f_o$. The arrow indicates the change with increasing temperature.}
 \label{fig.03}
\end{figure}

Darcy velocities of every component in both phases (neglecting mass diffusion) are
\begin{equation}
u_{o_i} = {X}_i u_o,
\quad
u_{g_i} = Y_i u_g,
\quad
u_{g_k} = Y_k u_g,
\quad
u_{g_r} = Y_r u_g,
\label{eqM.10}
\end{equation}
where $u_o$ and $u_g$ are Darcy velocities of the oleic and gas phase. With the unknown total Darcy velocity $u = u_o+u_g$, we express the phase velocities as
\begin{equation}
u_o = u f_o,
\quad
u_g = u-u_o = u (1-f_o).
\label{eqM.11}
\end{equation}
Here $f_o(T,\mathbf{X},s_o)$ is the oil fractional flow function, which is a given function of the dependent variables. As a function of oil saturation, $f_o$ has a characteristic S-shape (see \cref{fig.03}): it is nondecreasing with a single inflection point and satisfies the limiting conditions
\begin{equation}
f_o = \partial f_o/\partial s_o = 0 \textrm{  at  } s_o = 0 \textrm{ or } 1.
\label{eqA1}
\end{equation}
Mobility of oil increases with temperature, which yields the condition
\begin{equation}
\frac{\partial f_o}{\partial T} \ge 0.
\label{eqA1bbb}
\end{equation}
In \cref{sec:numer} we provide a specific example, but we consider a general form of $f_o$ in the theoretical part.

We use the ideal gas law to define
\begin{equation}
\rho_g = P/RT.
\label{eqM.IG}
\end{equation}
Though pressure gradient is nonzero, pressure variations within a relatively short combustion wave can be neglected. So we consider a constant pressure $P$ for physical properties of the phases.

For the source terms $W_{r_i}$ and $W_{v_i}$, it is important to formulate the equilibrium conditions,
\begin{align}
W_{r_i} &= 0: \ \   Y_k = 0 \textrm{ \ or \ } X_is_o = 0;
\label{eqWReq}
\\[5pt]
W_{v_i} &= 0: \ \   Y_i = Y_i^{eq} \textrm{ \ or \ } [X_is_o = 0 \textrm{ and } Y_i \le Y_i^{eq}].
\label{eqWReqB}
\end{align}
Equation \cref{eqWReq} implies no reaction in the absence of one of the reactants.
Conditions \cref{eqWReqB} mean no phase transition in the equilibrium state with a given function $Y_i^{eq}(T,\mathbf{X})$, or in a dry reservoir.
Equilibrium fractions $Y_i^{eq}$ increase with temperature
\begin{equation}
\frac{\partial Y_i^{eq}}{\partial T} > 0 \ \ \ i=1,\ldots,N.
\label{eq.dyeq}
\end{equation}
The boiling temperature $T = T_b(\mathbf{X})$ is defined by the condition
$Y_1^{eq}+\ldots+Y_N^{eq} = 1$.
Out of equilibrium, the signs of these rates are
\begin{eqnarray}
W_{r_i} > 0,\quad
\mathrm{sgn}\, W_{v_i} =
\mathrm{sgn}\, (Y_i^{eq}-Y_i).
\label{eqWReq2}
\end{eqnarray}
In our analysis we assume that the reaction rates $W_{r_i}$ and $W_{v_i}$ are very large when the system is out of equilibrium (relaxation times are very small). In this case their specific functional form will not be important.
For reaction rate, this implies that the oxygen reacts completely at large temperatures.

\subsection{Balance laws for total oleic and gas phases}

Taking the sum of Eqs.~\cref{eqM.xi} (divided by $\rho_i$) for all $N$ components and using \cref{eqM.10}, \cref{eqM.11} with $X_1+\cdots+X_N = 1$, we obtain the balance law for the total liquid phase as
\begin{equation}
 \frac{\partial}{\partial t}\varphi s_o
 +
 \frac{\partial}{\partial x} u f_o
 =
 -
 \sum_{i=1}^N
 \frac{\nu_{o_i}}{\rho_i}W_{r_i}
 -
 \sum_{i=1}^N
 \frac{1}{\rho_i}W_{v_i}.
 \label{eqM.so}
\end{equation}
Similarly, the sum of Eqs.~\cref{eqM.yi}--\cref{eqM.yr} yields the balance law for the total gas phase as
\begin{equation}
\frac{\partial}{\partial t}\varphi \rho_g s_g
+
\frac{\partial}{\partial x}\rho_g (1-f_o) u
=
\sum_{i=1}^N (\nu_{g_i} - 1)W_{r_i} + \sum_{i=1}^N W_{v_i}.
\label{eqM.sg}
\end{equation}

\subsection{Dimensionless equations}
\label{subsec:nondim}

In order to render the governing equations dimensionless, we introduce the ratios
\begin{equation}
\tilde{t}      = \frac{t}{t^*},\quad
\tilde{x}      = \frac{x}{x^*},\quad
\theta         = \frac{T-T_{ini}}{\Delta T^*},\quad
\tilde{u}      = \frac{u}{\varphi v^*},\quad
\tilde{\rho}_g = \frac{\rho_g}{\rho_g^*},
\label{eqN.1}
\end{equation}
where the reference quantities denoted by an asterisk are
\begin{equation}
t^* = \frac{x^*}{v^*},\quad
x^* = \frac{\lambda}{(1-\varphi)C_m v^*},\quad
v^* = \frac{Q_{r_1}\rho_g^*u^{inj}Y_k^{inj}}{(1-\varphi)C_m\Delta T^*},\quad
\rho_g^* = \frac{P}{RT_{ini}},\quad
\Delta T^* = T^*-T_{ini},
\label{eqN.2}
\end{equation}
and $T^*$ is some characteristic value, e.g., the boiling temperature for the initial liquid phase.
The dimensionless variable $\theta$ describes the temperature distribution
relative to the initial condition ($\theta = 0$ at $T = T_{ini}$).
The reference quantities $t^*$, $x^*$ and $v^*$ are obtained by considering a wave whose combustion heat raises the rock temperature to $T^*$, with the injected air at Darcy speed $u^{inj}$ and oxygen fraction $Y_k^{inj}$.

Dimensionless parameters are introduced as
\begin{equation}
\alpha_o    = \frac{\varphi C_o}{(1-\varphi)C_m},\ \
\alpha_g    = \frac{\varphi c_g\rho_g^*}{(1-\varphi)C_m},\ \
\beta_i     = \frac{\rho_g^*}{\rho_i},\ \
\gamma_{v_i} = \frac{Q_{v_i}}{Q_{r_1}},\ \
\gamma_{r_i} = \frac{Q_{r_i}}{Q_{r_1}},\ \
\theta_0 = \frac{T_{ini}}{\Delta T^*},\ \
\sigma = \frac{\varphi v^*}{ u^{inj}}.
\label{eqNS.6}
\end{equation}
We will adopt some extra simplifications, which considerably reduce complexity of the equations without affecting much the solution: we set $\gamma_{v_i} = 0$ by neglecting the heat of vaporization (subdominant to the heat of combustion), take  $\gamma_{r_i} = 1$ by neglecting a difference in the combustion heat for different components (taken per mole of oxygen), and put $\nu_{g_i} = 1$ implying no net gas production in the reaction. Some parameters can be considered as small due to different physical reasons, which include $\alpha_g$ (small heat capacity of gas compared to the rock matrix), $\beta_i$ (small gas density), $\sigma$ (small combustion wave speed compared to gas speed), and $\nu_{o_i}$ (scission of large hydrocarbon moleculas).

In the governing system \cref{eqM.xi}--\cref{eqM.t}, it is convenient to use \cref{eqM.so} and \cref{eqM.sg} instead of \cref{eqM.xi} for $X_N$ and \cref{eqM.yr}.
The resulting system written in dimensionless variables becomes (we drop the tildes for simplicity)
\begin{align}
 \frac{\partial X_i s_o}{\partial t}
 +
 \frac{\partial X_i u f_o}{\partial x}
 &=
 -\nu_{o_i}\beta_i w_{r_i} - \beta_i w_{v_i}, \ \ \ i=1,\ldots,N-1,
 \label{eqND.Xi}
 \\[5pt]
 \frac{\partial s_o}{\partial t}
 +
 \frac{\partial u f_o}{\partial x}
& =
 -
 \sum_{i=1}^N \nu_{o_i}\beta_i w_{r_i}
 -
 \sum_{i=1}^N \beta_i w_{v_i},
\label{eqND.002}
\\[5pt]
 \frac{\partial Y_i S_g}{\partial t}
 +
 \frac{\partial u F_g Y_i}{\partial x}
& =
 w_{v_i},  \ \ \ i=1,\ldots,N,
\label{eqND.004}
\\[5pt]
 \frac{\partial Y_k S_g}{\partial t}
 +
 \frac{\partial u F_g Y_k}{\partial x}
&=
 - \sum_{i=1}^N w_{r_i},
\label{eqND.005}
\\[5pt]
 \frac{\partial S_g}{\partial t}
 +
 \frac{\partial u F_g}{\partial x}
&= \sum_{i=1}^N w_{v_i},
\label{eqND.003}
\end{align}
\begin{equation}
 \frac{\partial}{\partial t}\left(1 + \alpha_o s_o + \alpha_g S_g\right)\theta
 +
 \frac{\partial}{\partial x}\left(\alpha_o f_o + \alpha_g F_g\right)u\theta
=
 \frac{\partial^2 \theta}{\partial x^2}
 +
 \frac{\sigma}{Y_k^{inj}}
 \sum_{i=1}^N w_{r_i},
\label{eqND.001}
\end{equation}
where $w_{r_i}(\theta,\mathbf{X},Y_k,s_o)$ and $w_{v_i}(\theta,\mathbf{X},\mathbf{Y},s_o)$ are the dimensionless reaction rates. Dimensionless gas density and
the temperature-corrected gas saturation and flux are defined as
\begin{equation}
 \rho_g = \frac{1}{1+\theta/\theta_0},
 \quad
 S_g = (1 - s_o)\rho_g,
 \quad
 F_g = (1 - f_o)\rho_g.
 \label{eqND.007}
\end{equation}
The fractional flow function and equilibrium gas fraction are now expressed as $f_o(\theta,\mathbf{X},s_o)$ and $Y_i^{eq}(\theta,\mathbf{X})$,
where the $N$ functions $Y_i^{eq}(\theta,\mathbf{X})$ have the property
\begin{equation}
\frac{\partial Y_i^{eq}(\theta,\mathbf{X})}{\partial \theta} > 0.
\label{eqND.Yheq}
\end{equation}

\section{Combustion wave}
\label{sec:CC}

A typical problem for oil recovery is formulated under specific boundary conditions at $x = 0$ (modeling an injection well) and initial reservoir conditions for $x > 0$. Usually a full solution is only accessible by numerical simulations. However, at large times, the solution may be represented asymptotically as a sequence of waves, which can be studied analytically. Our interest in this paper is the combustion wave, where oxygen reacts with oil components. We assume that the combustion wave has a stationary profile traveling at a constant speed $v$, where all variables depend on a single traveling coordinate $\xi = x - v t$. In this case, system \cref{eqND.Xi}--\cref{eqND.001} reduces to the ordinary differential equations
\begin{align}
\frac{d X_i \psi_o}{d \xi}
&=
- \nu_{o_i}\beta_i w_{r_i} - \beta_i w_{v_i}, \quad i=1,\ldots,N-1,
\label{eqCW.Xi}
\\[5pt]
\frac{d \psi_o}{d \xi}
&=
- \sum_{i=1}^N \nu_{o_i}\beta_i w_{r_i} - \sum_{i=1}^N \beta_i w_{v_i},
\label{eqCW.psio}
\\[5pt]
\frac{d Y_i \psi_g}{d \xi}
&=
w_{v_i}, \quad i = 1,\ldots,N,
\label{eqCW.Yi}
\\[5pt]
\frac{d Y_k \psi_g}{d \xi}
&=
-
\sum_{i=1}^N w_{r_i},
\label{eqCW.Yk}
\\[5pt]
\frac{d \psi_g}{d \xi}
&=
\sum_{i=1}^N w_{v_i},
\label{eqCW.psig}
\\[5pt]
\frac{d}{d \xi}
\left[\left(- v + \alpha_o \psi_o + \alpha_g \psi_g\right)\theta\right]
&=
\frac{d^2 \theta}{d \xi^2}
+
\frac{\sigma}{Y_k^{inj}}
\sum_{i=1}^N w_{r_i},
\label{eqCW.t}
\end{align}
where we denoted the liquid and gas fluxes in the moving reference frame as
\begin{equation}
 \psi_o = u f_o - v s_o,
 \quad
 \psi_g = u F_g - v S_g = \rho_g(u-v-\psi_o).
 \label{eqCW.007}
\end{equation}
All the dependent variables are now functions of $\xi$ only.

Upstream and downstream states of the combustion wave must be at equilibrium. This means that no oil remains on the upstream side, where air is injected, and no oxygen reaches the downstream side. The oleic and gas phases are in thermodynamic equilibrium on the downstream side. As shown in~\cite{mailybaev2011resonance}, such a combustion wave is followed by the slower thermal wave on the upstream side, characterized by the change of temperature between the cold injected gas and rock heated by combustion. On the downstream side, follows the faster Buckley--Leverett wave, where oil saturation changes to its value in the initial reservoir, with no alteration in oil composition or temperature, \cref{fig.03}.

Let us specify the limiting constant states of the combustion wave. On the upstream (hot) side of the wave, the temperature $\theta^u > 0$ is unknown, all oil is displaced and/or consumed, the gaseous phase contains injected air with oxygen fraction $Y_k  = Y_k^{inj}$ and no oil components.
A more subtle condition is related to the upstream dimensionless value $\tilde{u}$ of the Darcy velocity in \cref{eqN.1}. For cold injected air with dimensional Darcy speed $u^{inj}$, expressions \cref{eqN.1} and \cref{eqNS.6} yield $\tilde{u} = 1/\sigma$. This value must be temperature-corrected due to the dependence of gas density $\rho_g$ on temperature, see \cref{eqND.007}, providing the value $\tilde{u} = (1 + \theta^u / \theta_0)/\sigma$. In fact, this expression still requires some extra correction depending on the speed of thermal wave, but this correction is small and can be neglected, see e.g.~\cite{mailybaev2013recovery}.
Therefore, we have the upstream limiting state (again dropping the tilde) as
\begin{equation}
 \xi\rightarrow-\infty: \quad
 \theta    = \theta^u, \quad
 Y_i       = 0, \quad
 Y_k       = Y_k^{inj}, \quad
 s_o       = 0, \quad
 u         = \frac{1 + \theta^u / \theta_0}{\sigma}.
 \label{eqCW.009}
\end{equation}
Since $s_o = 0$ implies $f_o = 0$ by \cref{eqA1}, the upstream values of the fluxes \cref{eqCW.007} are given by
\begin{equation}
 \xi\rightarrow-\infty: \quad
 \psi_o   = 0, \quad
 \psi_g   = \rho_g(u-v) \approx \rho_g u = \frac{1}{\sigma},
 \label{eqCW.010}
\end{equation}
where we neglected the much smaller combustion wave speed $v$ compared to the gas speed $u$. Note that $s_o = 0$ also means that the composition vector $\mathbf{X}$ is irrelevant at this limiting state.

On the downstream (cold) side of the wave, the oil is at its initial temperature $\theta=0$ and composition $X_i = X_i^{ini}$.
We assume complete oxygen consumption at highest temperatures of the combustion wave, which happens for low and moderate injection speeds~\cite{santos2016}.
Together with the thermodynamic equilibrium condition, this yields the following downstream limiting state
\begin{equation}
 \xi\rightarrow+\infty:	 \quad
 \theta = 0, \quad
 X_i = {X}_i^{ini},\quad
 Y_i    = Y_i^{ini}, \quad
 Y_k    = 0, \quad
 s_o    = s_o^d, \quad
 u      = u^d,
 \label{eqCW.011}
\end{equation}
where $Y_i^{ini} = Y_i^{eq}(0,\mathbf{X}^{ini})$, while the oil saturation $s_o^d$ and gas speed $u^d$ are unknown. For the downstream fluxes, $\psi_o = \psi_o^d$ and $\psi_g = \psi_g^d$, expressions \cref{eqCW.007} and \cref{eqCW.011} yield
\begin{equation}
 \xi\rightarrow+\infty:	 \quad
\psi_o^d = u^df_o(0,\mathbf{X}^{ini},s_o^d)-vs_o^d,\quad
\psi_g^d = u^d-v-\psi_o^d.
 \label{eqCW.011fluxes}
\end{equation}

In summary, the unknowns of the limiting states are $\theta^u$, $s_o^d$, $u^d$ and the wave velocity $v$. 
The theory that provides these quantities by solving equations \cref{eqCW.Xi}--\cref{eqCW.t} with the boundary conditions \cref{eqCW.009}--\cref{eqCW.011fluxes} is the main goal of this paper. In the next section we show that three relations for these four unknowns can be determined from the overall balance relations analogous to the Rankine--Hugoniot conditions. But finding the remaining condition, namely, a condition for the wave speed $v$, requires sophisticated analysis of a singularity at an internal point of the wave profile.

\section{Conservation laws}
\label{sec:Conl}

Equations \cref{eqCW.Xi} and \cref{eqCW.Yi} can be used to express the source terms as
\begin{equation}
 w_{v_i} = \frac{d Y_i\psi_g}{d\xi},
 \quad
 w_{r_i} = -\frac{1}{\nu_{o_i}\beta_i}\frac{d X_i\psi_o}{d\xi}
 -\frac{1}{\nu_{o_i}}
\frac{d Y_i\psi_g}{d\xi}.
 \label{eqCW.w}
\end{equation}
We can then obtain source-free equations (conservation laws) by substituting \cref{eqCW.w} into Eqs.~\cref{eqCW.Yk} and \cref{eqCW.psig}. This yields
\begin{align}
 \frac{d}{d \xi}
 \left(
 Y_k \psi_g
 -
\psi_o \sum_{i=1}^N\frac{X_i}{\nu_{o_i}\beta_i}
 -
\psi_g \sum_{i=1}^N\frac{Y_i}{\nu_{o_i}}
 \right)
 &=
 0,
 \label{eqSF.yk}
 \\[5pt]
 \frac{d}{d\xi}
 \left(\psi_g-\psi_g\sum_{i=1}^N Y_i\right)
 &=
 0,
 \label{eqSF.psig}
\end{align}
Similarly, substituting \cref{eqCW.Yk} into \cref{eqCW.t}, we obtain
\begin{equation}
 \frac{d}{d\xi}
 \left(
 (-v + \alpha_o\psi_o + \alpha_g\psi_g)\theta - \frac{d\theta}{d\xi} + \sigma \psi_g\frac{Y_k}{Y_k^{inj}}
 \right) = 0.
 \label{eqSF.t}
\end{equation}

Expressions in the parentheses of \cref{eqSF.yk}--\cref{eqSF.t} determine three conserved quantities along the wave profile. We can compare their values at the limiting states as
\begin{align}
 \frac{Y_k^{inj}}{\sigma}
 &=
 -
 \psi_o^d\sum_{i=1}^N\frac{X_i^{ini}}{\nu_{o_i}\beta_i}
 -
 \psi_g^d\sum_{i=1}^N\frac{Y_i^{ini}}{\nu_{o_i}},
 \label{eq.yk}
 \\[5pt]
 \frac{1}{\sigma}
 &=
 \psi_g^d\left(1 - \sum_{i=1}^N Y_i^{ini}\right),
 \label{eq.psig}
 \\[5pt]
 \left(-v+\frac{\alpha_g}{\sigma}\right)\theta^u + 1
 &=
 0,
 \label{eq.tu}
\end{align}
where the terms on the left correspond to the upstream side \cref{eqCW.009} and \cref{eqCW.010},
while the terms on the right correspond to the downstream side \cref{eqCW.011}.
Here we used the fact that $d\theta/d\xi=0$ at the limiting constant states.

Equation \cref{eq.tu} determines the upstream temperature as
\begin{equation}
 \theta^u = \frac{1}{v-\alpha_g/\sigma},
 \label{eq.tuF}
\end{equation}
while Eqs.~\cref{eq.yk} and \cref{eq.psig} give the downstream fluxes as
\begin{equation}
 \psi_o^d =
 - \left(\frac{Y_k^{inj}}{\sigma}
 +\psi_g^d
 \sum_{i=1}^N \frac{Y_i^{ini}}{\nu_{o_i}}
\right)
 \left(\sum_{i=1}^N \frac{X_i^{ini}}{\nu_{o_i}\beta_i}\right)^{-1},
\quad
 \psi_g^d = \frac{1}{\sigma}\left(1 - \sum_{i=1}^N Y_i^{ini}\right)^{-1}.
 \label{eq.psiodA}
\end{equation}

Equations \cref{eq.tuF} and \cref{eq.psiodA} with the fluxes expressed from \cref{eqCW.011fluxes} provide three equations for the four unknowns $\theta^u$, $s_o^d$ and $u_g^d$ and $v$. We conclude that the system of determining relations provided by the total balance is incomplete, and we have to analyze the wave profile for determining the wave parameters.

\section{Wave profile equations}
\label{sec:wprof}

In this section, we derive equations for the wave profile.
We will obtain expressions for the fluxes $\psi_o$ and $\psi_g$ and express the governing equations for the $2N+1$ variables, $\theta$, $\mathbf{X}$, $\mathbf{Y}$ and $Y_k$, in terms of these fluxes.
The conserved quantities \cref{eqSF.yk} and \cref{eqSF.psig} at an arbitrary point $\xi$ are
\begin{align}
 Y_k \psi_g
 -
\psi_o \sum_{i=1}^N\frac{X_i}{\nu_{o_i}\beta_i}
 -
\psi_g \sum_{i=1}^N\frac{Y_i}{\nu_{o_i}}
 &=
 \frac{Y_k^{inj}}{\sigma},
 \label{ExeqSF.yk}
 \\[5pt]
\psi_g-\psi_g\sum_{i=1}^N Y_i
 &=
 \frac{1}{\sigma},
 \label{ExeqSF.psig}
\end{align}
where the values on the right-hand sides are taken from \cref{eq.yk} and \cref{eq.psig}.
Solving these equation for $\psi_o$ and $\psi_g$ yields
\begin{equation}
 \psi_o =
 - \left(\frac{Y_k^{inj}}{\sigma}
  -\psi_gY_k+\psi_g
 \sum_{i=1}^N \frac{Y_i}{\nu_{o_i}}
\right)
 \left(\sum_{i=1}^N \frac{X_i}{\nu_{o_i}\beta_i}\right)^{-1},
  \quad
 \psi_g =
 \frac{1}{\sigma}\left(1-\sum_{i=1}^N Y_i\right)^{-1}.
 \label{NeqFOLD.psio}
\end{equation}
Expressing now $u$ and $f_0$ from \cref{eqCW.007}, one obtains
\begin{align}
u  &= \frac{\psi_g}{\rho_g}+\psi_o  +v,
 \label{eqCW.007BB}
\\
f_o &= \frac{v s_o}{u}
+\frac{\psi_o}{u}.
 \label{eqCW.007B}
\end{align}
These two expressions, where $f_o(\theta,\mathbf{X},s_o)$ is a given function and $\psi_o$, $\psi_g$ are determined by \cref{NeqFOLD.psio}, provide the gas velocity $u$ (explicitly) and the oil saturation $s_o$ (implicitly) in terms of other variables.

Analogous integration of Eq.~\cref{eqSF.t} yields the ordinary differential equation
\begin{equation}
 \frac{d\theta}{d\xi} =
(-v + \alpha_o\psi_o + \alpha_g\psi_g)\theta + \sigma \psi_g\frac{Y_k}{Y_k^{inj}}.
\label{ExtraA1}
\end{equation}
The remaining equations of system \cref{eqCW.Xi}--\cref{eqCW.t}, which cannot be integrated explicitly, are
\begin{align}
\frac{d X_i \psi_o}{d \xi}
&=
- \nu_{o_i}\beta_i w_{r_i} - \beta_i w_{v_i}, \quad i=1,\ldots,N-1,
\label{NeqCW.Xi}
\\[5pt]
\frac{d Y_i \psi_g}{d \xi}
&=
w_{v_i}, \quad i = 1,\ldots,N,
\label{NeqCW.Yi}
\\[5pt]
\frac{d Y_k \psi_g}{d \xi}
&=
-
\sum_{i=1}^N w_{r_i}.
\label{NeqCW.Yk}
\end{align}

\subsection{Overall structure and the fold singularity}
\label{subsec:overall}

With $\psi_o$, $\psi_g$ and $u$ substituted from \cref{NeqFOLD.psio} and \cref{eqCW.007BB}, we obtain the system of $2N+2$ equations \cref{eqCW.007B}--\cref{NeqCW.Yk} dependent on $2N+2$ variables: $\mathbf{x} = (\theta,\mathbf{X},\mathbf{Y},Y_k)\in \mathbb{R}^{2N+1}$ and $s_0$. The left-hand sides in \cref{NeqCW.Xi}--\cref{NeqCW.Yk} are not resolved with respect to derivatives. Thus, we can write the structure of this system as
\begin{align}
H(\mathbf{x},s_o) &= 0,
\label{ExtraA2}
\\[5pt]
\mathbf{A}(\mathbf{x},s_0)\dot{\mathbf{x}} &= \mathbf{b}(\mathbf{x},s_0).
\label{ExtraA2bb}
\end{align}
Here $H$ is a scalar function corresponding to \cref{eqCW.007B}, while the $(2N+1)\times(2N+1)$ matrix $\mathbf{A}$ and the vector $\mathbf{b}\in\mathbb{R}^{2N+1}$ represent \cref{ExtraA1}--\cref{NeqCW.Yk}.
It is instructive to write the full derivative of the first equation in the form
\begin{equation}
\frac{\partial H}{\partial \mathbf{x}}\,\dot{\mathbf{x}}
+\frac{\partial H}{\partial s_0}\,\dot{s}_0
= 0.
\label{ExtraA3}
\end{equation}
We see that the wave profile in the space $(\mathbf{x},s_o) \in \mathbb{R}^{2N+2}$ is determined by the vector field defined implicitly by the $2N+2$ equations \cref{ExtraA2bb}--\cref{ExtraA3}, and limited to the integral hypersurface given by equation \cref{ExtraA2}.

This implicit formulation for the vector field \cref{ExtraA2bb}--\cref{ExtraA3} has intrinsic singularities, when the system Jacobian with respect to time-derivatives $\dot{\mathbf{x}}$ and $\dot{s}_o$ vanishes. At such points, the values of $\dot{\mathbf{x}}$ and $\dot{s}_o$ are either undetermined or nonunique. We can classify these singularities into two classes. The first is given by the points, where $\det\mathbf{A} = 0$ and equations \cref{ExtraA2bb} cannot be resolved uniquely with respect to $\dot{\mathbf{x}}$. The second type is related to the condition
\begin{equation}
\frac{\partial H}{\partial s_0} = 0,
\label{ExtraA4}
\end{equation}
when the derivative $\dot{s}_o$ in \cref{ExtraA3} remains undefined. This second type corresponds to the folding singularity of hypersurface \cref{ExtraA2}  when projected onto the space $\mathbf{x}$ along the axis $s_o$.

Expressing $H = uf_o-vs_o-\psi_o$ from \cref{eqCW.007B} with formulae \cref{NeqFOLD.psio}--\cref{eqCW.007BB}, we write condition \cref{ExtraA4} for our system in the form
\begin{equation}
\textrm{fold condition (resonance):}\quad
v = u\,\frac{\partial f_o}{\partial s_o}.
\label{ExtraA4B}
\end{equation}
The right-hand side of this expression represents the characteristic speed of the Buckley--Leverett wave~\cite{Smoller1983}. For this reason, condition \cref{ExtraA4B} was termed in~\cite{mailybaev2011resonance} as the resonance condition.

We now argue that the combustion wave profile passes through the fold. Indeed, at the upstream state \cref{eqCW.009}, we have $s_o = 0$. Hence, $\partial f_o/\partial s_o = 0$ according to \cref{eqA1} and
\begin{equation}
\xi \to -\infty:\quad
v  > u\,\frac{\partial f_o}{\partial s_o} = 0.
\label{ExtraA4Bup}
\end{equation}
With the wave sequence solution described in \cref{sec:CC}, there is a Buckley--Leverett (saturation) wave traveling faster than the combustion wave. In this wave the oil saturation is adjusted to meet the initial reservoir conditions~\cite{mailybaev2011resonance}. Since $\lambda_{BL} = u \,(\partial f_o/\partial s_o)$ is the characteristic speed of the Buckley--Leverett wave, the described wave sequence requires $v < \lambda_{BL}$ at the downstream state, i.e.,
\begin{equation}
\xi \to +\infty:\quad
v  < u\,\frac{\partial f_o}{\partial s_o}.
\label{ExtraA4Bdown}
\end{equation}
Properties \cref{ExtraA4Bup} and \cref{ExtraA4Bdown} show that the fold condition \cref{ExtraA4B} must be satisfied at some internal point of the wave profile.
 We will see that this fact is crucial for determining the wave parameters.

\subsection{Small relaxation-time asymptotic}
\label{subsec:asym}

Due to complexity and high dimension of system \cref{ExtraA2}--\cref{ExtraA2bb}, analysis of the wave profile in the general case is a very difficult task. However, one physical aspect of the problem makes it feasible. As we mentioned in \cref{sec2.2}, relaxation times for vaporization and reaction are very small for our problem. This can be used to derive an asymptotic form of the equations.

Formally, the asymptotic equations can be derived by introducing small parameters, $\varepsilon_{r_i} \ll 1$ and $\varepsilon_{v_i} \ll 1$, that represent the relaxation times of reaction and vaporization. In general, these parameters define a set of different scales, but for simplicity we will restrict our analysis to a single dominant small scale, $\varepsilon = \max\{\varepsilon_r,\varepsilon_v\} \ll 1$.  In the equations, this small parameter is introduced by the formal substitution
\begin{equation}
w_{r_i} \mapsto \frac{\hat{w}_{r_i}}{\varepsilon}, \quad
w_{v_i} \mapsto \frac{\hat{w}_{v_i}}{\varepsilon},
\label{ExtraA5}
\end{equation}
for the reaction and vaporization rates. This substitution represents a singular perturbation for Eqs.~\cref{NeqCW.Xi}--\cref{NeqCW.Yk}, which divides the wave profile into equilibrium and out-of-equilibrium regions, see e.g. \cite{jones1995geometric}.
In the context of geometric singular perturbations, they correspond to the fast (out-of-equilibrium) and slow (equilibrium) regions.
Equations \cref{NeqCW.Xi}--\cref{NeqCW.Yk} then, become
\begin{align}
\frac{d X_i \psi_o}{d \xi}
&=
- \nu_{o_i}\beta_i \frac{\hat{w}_{r_i}}{\varepsilon} - \beta_i \frac{\hat{w}_{v_i}}{\varepsilon}, \quad i=1,\ldots,N-1,
\label{NeqCW.XiB}
\\[5pt]
\frac{d Y_i \psi_g}{d \xi}
&=
\frac{\hat{w}_{v_i}}{\varepsilon}, \quad i = 1,\ldots,N,
\label{NeqCW.YiB}
\\[5pt]
\frac{d Y_k \psi_g}{d \xi}
&=
-
\sum_{i=1}^N \frac{\hat{w}_{r_i}}{\varepsilon},
\label{NeqCW.YkB}
\end{align}
which are valid on the hypersurface
\begin{equation}
f_o = \frac{v s_o}{u} + \frac{\psi_o}{u}.
 \label{eqCW.007BC}
\end{equation}

In the equilibrium region, one multiplies both sides of the equations by a small parameter and takes the limit, e.g.,
\begin{equation}
\varepsilon\,\frac{d Y_i \psi_g}{d \xi}
= \hat{w}_{v_i}\ \xrightarrow{\varepsilon \to 0}\ \hat{w}_{v_i} = 0.
\label{ExtraA6}
\end{equation}
Hence, in the equilibrium region, one arrives asymptotically to the system
\begin{equation}
\text{equilibrium region}:\quad
w_{r_i} \approx 0,\quad  w_{v_i} \approx 0.
\label{ExtraA7}
\end{equation}
Equations for the out-of-equilibrium region are constructed by introducing the new spacial scale $\hat\xi = \xi/\varepsilon$, describing a very thin region to which the reaction and vaporization are confined. In this case, equations \cref{NeqCW.Xi}--\cref{NeqCW.Yk} remain the same (with $\hat\xi$ instead of $\xi$), but Eq.~\cref{ExtraA1} reduces to $d\theta/d\hat\xi = 0$ in the limit $\varepsilon \to 0$. This implies that the change of temperature is negligible in a thin out-of-equilibrium region:
\begin{equation}
\text{out-of-equilibrium region}:\quad
\theta \approx const.
\label{ExtraA8}
\end{equation}

\section{Single component case}
\label{sec.ONE}

We start with the simplest case of a single-component, $N = 1$. Note that this case was considered earlier in~\cite{mailybaev2011resonance}, where additionally the thermal diffusion was neglected. This latter simplification decreased the system dimension allowing for the reduction to a planar implicit ordinary differential equations. In this section, we provide a more general view to this problem that will be generalized in Section~\ref{sec.Ncomp} to the general case of $N$ components.

In the case of a single component, one has $X_1 = 1$ and we will drop the component index ``1'' to simplify the notations.
From \cref{NeqFOLD.psio} and \cref{eqCW.007BB} with elementary manipulations we get
\begin{equation}
 \psi_o
 =
- \frac{\nu_{o}\beta(Y_k^{inj}-Y_k)  + \beta Y(1-\nu_oY_k^{inj})}{\sigma\left(1-Y\right)},\quad
 \psi_g
 =
 \frac{1}{\sigma\left(1-Y\right)},
 \quad
 u  = \frac{\psi_g}{\rho_g}+\psi_o  +v.
 \label{eqONE.psi}
\end{equation}
Then, Eqs.~\cref{eqCW.007B}--\cref{NeqCW.Yk} for a vector field on a hypersurface in the 4-dimensional space $(\theta,Y,Y_k,s_o)$ take the form
\begin{align}
 f_o(\theta,s_o)
 &=
\frac{v s_o}{u}
+\frac{\psi_o}{u},
 \label{ExtraA9}
\\[5pt]
 \frac{d\theta}{d\xi}
 &=
 (-v + \alpha_o\psi_o + \alpha_g\psi_g)\theta + \sigma \psi_g\frac{Y_k}{Y_k^{inj}},
 \label{eqONE.t}
 \\[5pt]
\frac{d Y\psi_g}{d \xi}
 &=
w_v,
 \label{eqONE.Yi}
 \\[5pt]
 \frac{d Y_k\psi_g}{d \xi}
 &=
 -w_{r}.
 \label{eqONE.Ykw}
\end{align}
The last two equations can be modified using $\psi_g$ from \cref{eqONE.psi} as
\begin{align}
\frac{d Y}{d \xi}
 &=
\sigma (1-Y)^2w_v,
 \label{eqONE.YiNew}
 \\[5pt]
 \frac{d Y_k}{d \xi}
 &=
 -\sigma (1-Y)w_{r}-Y_kw_v.
 \label{eqONE.YkwNew}
\end{align}

The wave profile is a solution of three ordinary differential equations \cref{eqONE.t}, \cref{eqONE.YiNew} and \cref{eqONE.YkwNew} on a hypersurface in space $(\theta,Y,Y_k,s_o)$ given by Eq.~\cref{ExtraA9}. This solution must be a heteroclinic orbit connecting the stationary limiting states (see \cref{sec:CC}) as
\begin{equation}
 \xi\rightarrow-\infty: \quad
 \theta =  \theta^u, \quad
 Y      = 0, \quad
 Y_k       = Y_k^{inj}, \quad
 s_o       = 0,
 \label{ExtraS1}
\end{equation}
\begin{equation}
 \xi\rightarrow+\infty:	 \quad
 \theta = 0, \quad
 Y   = Y^{ini}, \quad
 Y_k    = 0,\quad
  s_o       = s_o^d,
 \label{ExtraS1bb}
\end{equation}
with the unknown $\theta^u$, $s_o^d$ and  wave speed $v$.

\subsection{Folded hypersurface}
\label{sec.fold}

We start with the analysis of Eq.~\cref{ExtraA9}. As we mentioned in \cref{subsec:nondim}, the parameters $\beta$, $\sigma$ and $\nu_o$ are small. With this assumption, relations \cref{eqONE.psi} yield approximately
\begin{equation}
 \psi_o
 \approx
- \frac{\beta Y}{\sigma\left(1-Y\right)},
 \quad
 \psi_g
 =
 \frac{1}{\sigma\left(1-Y\right)},
\quad
 u  \approx
  \frac{1+\theta/\theta_0}{\sigma\left(1-Y\right)},
 \label{ExtraFS1}
\end{equation}
where we expressed $\rho_g$ from \cref{eqND.007}. 
This approximation is valid for any ratios among small parameters $\beta$, $\sigma$ and $\nu_o$, and it follows that $u \sim \psi_g$, $\psi_g \gg 1$ and $\psi_g \gg \psi_o$. 
Then Eq.~\cref{ExtraA9} with the folding condition \cref{ExtraA4B} take the form
\begin{equation}
 f_o(\theta,s_o)  = \frac{v(1-Y)}{1+\theta/\theta_0}\,\sigma s_o-\frac{\beta Y}{1+\theta/\theta_0},
 \quad
 \frac{\partial}{\partial s_o}f_o(\theta,s_o)  = \frac{v(1-Y)}{1+\theta/\theta_0}\,\sigma.
\label{ExtraFS2}
\end{equation}

With the S-shape form of the function $f_o$, see \cref{fig.03}, and small parameters $\sigma$ and $\beta$, one can see that the solution $s_o$ of the first equation is small. Let us approximate $f_o$ with properties \cref{eqA1} as
\begin{equation}
f_o = {\textstyle\frac{1}{2}}f_2(\theta)s_o^2+o(s_o^2),
\label{ExtraFS3}
\end{equation}
where $f_2(\theta) = \partial^2 f_o/\partial s_o^2$ is evaluated at $s_o = 0$. 
Neglecting a small higher-order term $o(s_o^2)$, after a lengthy but elementary derivation one solves \cref{ExtraFS2} and \cref{ExtraFS3} as
\begin{equation}
\textrm{fold}:\quad s_o = \frac{v(1-Y)}{f_2(\theta)(1+\theta/\theta_0)}\,\sigma, \quad
\frac{Y}{(1-Y)^2} = \frac{v^2\sigma^2}{2\beta f_2(\theta)(1+\theta/\theta_0)}.
\label{ExtraFS4}
\end{equation}
Recall that $f_2(\theta)$ is a nondecreasing positive function due to condition \cref{eqA1bbb}. Hence, the solution $Y = Y^f(\theta)$ of the second equation in \cref{ExtraFS4} is unique in the interval $0 \le Y < 1$, with a decreasing function $Y^f(\theta)$.

\begin{figure}
 \begin{center}
  \includegraphics[width=0.33\linewidth]{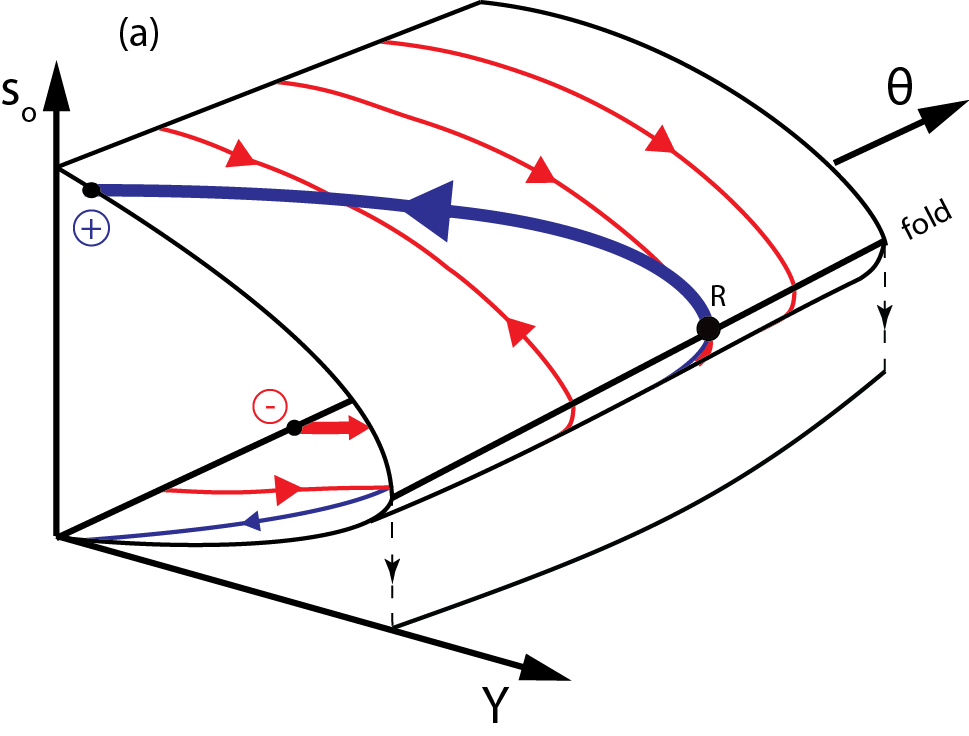}
  \hspace{3mm}
  \includegraphics[width=0.32\linewidth]{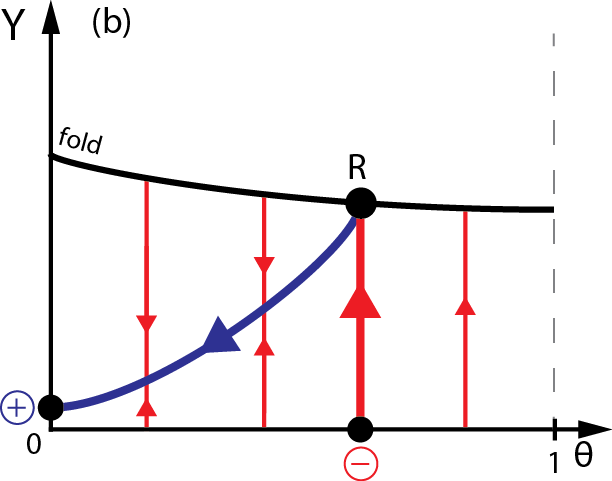}
  \hspace{3mm}
  \includegraphics[width=0.28\linewidth]{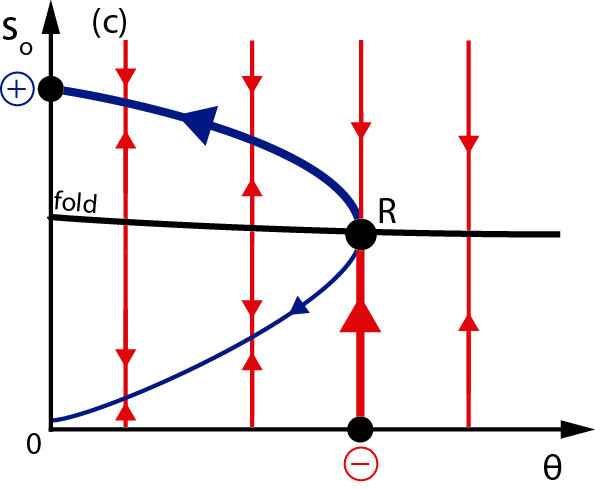}
  \end{center}
  \caption{(a) Folded integral surface in space $(\theta,Y,s_o)$ and its projection onto (b) the plane $(\theta,Y)$, (c) the plane $(\theta,s_o)$. The orbit corresponding to the equilibrium region is shown with a blue line, and the orbits in the out-of-equilibrium region by red lines. The bold line shows the wave profile connecting the two limiting states, from $\ominus$ to $\oplus$.}
 \label{fig.fold}
\end{figure}

The obtained results reveal the structure of the hypersurface \cref{ExtraA9} in the space $(\theta,Y,Y_k,s_o)$. In the first approximation, it does not depend on $Y_k$ and has a fold for small $s_o$. 
The geometry of this fold is shown schematically in \cref{fig.fold}.
We note that similar analysis can be done in a more realistic case, when $f_o = 0$ for saturations below a residual value, $s_o \le s_{or}$. In this case one substitutes $s_o^2$ by $(s_o-s_{or})^2$ in expression \cref{ExtraFS3} for $s_o > s_{or}$.

\subsection{Equilibrium and out-of-equilibrium regions}
\label{subsec:regions}

When the injected air comes into a contact with oil on the upstream hot side (see \cref{fig.03}), fast vaporization and reaction take place in a thin out-of-equilibrium region. 
The equilibrium region follows downstream, where no oxygen is left and oil components are close to the gas-liquid equilibrium. 
Since we are free to choose the origin for the coordinate $\xi$, we consider the out-of-equilibrium region for $\xi < 0$ and the equilibrium region for $\xi > 0$.

In the equilibrium region, conditions \cref{ExtraA7} yield
\begin{equation}
\textrm{equilibrium region:}\quad
Y = Y^{eq}(\theta),\quad
Y_k = 0,
\label{ExtraFS5}
\end{equation}
where $Y^{eq}(\theta)$ increases with temperature, see Eq.~\cref{eq.dyeq}.
Then the heat balance equation \cref{eqONE.t} with $\psi_o$, $\psi_g$ and $v$ expressed from \cref{ExtraFS1} and \cref{eq.tuF}  becomes
\begin{equation}
\frac{d\theta}{d\xi} =
- \left(\frac{1}{\theta^u} + \frac{(\alpha_o\beta-\alpha_g) Y}{\sigma(1-Y)}\right)\theta.
\label{ExtraFS6}
\end{equation}
Due to physical reasons (oil heat capacity per mole is larger than gas heat capacity~\cite{kokubun2016}),
we have $\alpha_o\beta > \alpha_g$, see \cref{eqNS.6}.
Thus, the temperature in \cref{ExtraFS6} decreases monotonously along the profile.

The results are demonstrated schematically in \cref{fig.fold}.
As we showed in \cref{sec.fold}, the fold line is given by a decreasing function $Y=Y^f(\theta)$, while $Y=Y^{eq}(\theta)$ increases
with $\theta$.
Therefore, the equilibrium part of the wave profile is transversal to the fold, when projected onto the plane $(\theta,Y)$.
Hence, the profile in the equilibrium region can only belong to one (upper) part of the hypersurface, extending at most to the fold line.

Condition \cref{ExtraA8} in the out-of-equilibrium region corresponds to $\xi < 0$, and the temperature
is found by matching with the limiting state \cref{ExtraS1}:
\begin{equation}
\textrm{out-of-equilibrium region:}\quad
\theta = \theta^u.
\label{ExtraFS7}
\end{equation}
As $Y = 0$ at the upstream limiting state, this region is characterized by vaporization, $w_v > 0$, see \cref{eqWReq2}.
Thus, according to \cref{eqONE.YiNew} and \cref{eqONE.YkwNew}, we have the monotonous increase of $Y$ and decrease of $Y_k$ with $\xi$.
The obtained results are summarized in \cref{fig.fold} by a family of red lines.
One can see that the out-of-equilibrium section of the wave profile is transversal to the fold when projected onto the plane $(\theta,Y)$, see \cref{fig.fold}(b).
Hence, the wave profile in this region belongs only to one (lower) part of the hypersurface, extending at most to the folding line.

\subsection{Resonance point}

Recall that the analysis in \cref{subsec:regions} is asymptotic, based on the assumption of fast reaction and vaporization. Therefore, the vector field on the folded surface presented in \cref{fig.fold} undergoes a rapid but continuous transition in a small neighborhood of the blue equilibrium line. A crucial conclusion of this asymptotic analysis is that the equilibrium and out-of-equilibrium regions on the surface lie at opposite sides of the fold line, i.e., this line separates these two regions. This result is robust to a change of parameters, at least for small perturbations, because it follows from qualitative properties of the underlying system (like folding and transversality).

At the folding line \cref{ExtraA4B}, both conditions \cref{ExtraFS5} and \cref{ExtraFS7} are satisfied, and we denote the corresponding value of oil saturation by $s_o = s_o^f$.
Combining these relations with \cref{ExtraA9}, the system of two equations
\begin{equation}
uf_o(\theta,s_o) =
v s_o+\psi_o,
\quad
v = u\,\frac{\partial f_o}{\partial s_o}
\label{ExtraFS8}
\end{equation}
is obtained at the resonance point
\begin{equation}
\theta = \theta^u,\ \
Y = Y^{eq}(\theta^u),\ \
Y_k = 0,\ \
s_o = s_o^f.
\label{ExtraFS8bb}
\end{equation}
Substituting $\theta^u$ from  \cref{eq.tuF} and $\psi_o$, $u$ from \cref{eqONE.psi},  equations \cref{ExtraFS8} define implicitly the wave speed $v$ and the oil saturation at the resonance point $s_o^f$. The same two equations were obtained when neglecting the heat conduction~\cite{mailybaev2011resonance}, where it was argued that the system possesses a unique solution. For practical reasons, these equations can be easily solved  numerically.

\begin{figure}
 \centering
 \includegraphics[width=0.5\linewidth]{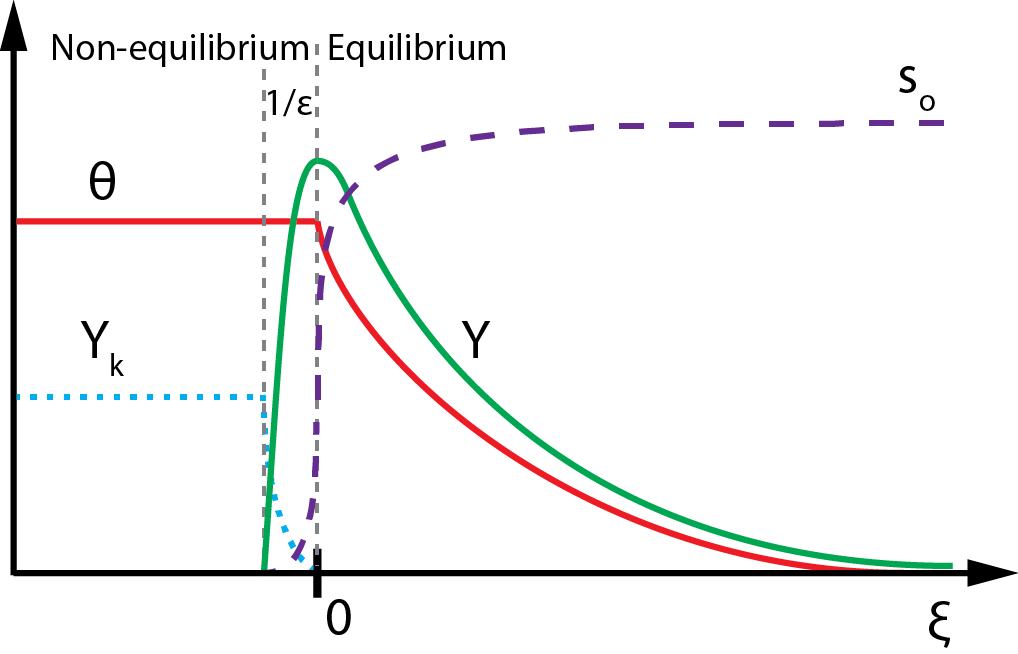}
 \caption{Wave profile with the thin non-equilibrium region and the large equilibrium region, separated by the resonance point $\xi = 0$. Shown are schematic graphs of the temperature $\theta$, oxygen fraction $Y_k$, gaseous hydrocarbon $Y$ and oil saturation $s_o$.}
 \label{fig.profile}
\end{figure}

When the wave speed $v$ is determined, other parameters of the combustion wave are found as explained in \cref{sec:Conl}. Namely, three equations in \cref{eq.tuF} and \cref{eq.psiodA} with the fluxes expressed from \cref{eqCW.011fluxes} can be solved with respect to the three unknowns: $\theta^u$, $s_o^d$ and $u_g^d$.
More detailed information on the wave profile can be obtained by constructing a solution in the equilibrium and out-of-equilibrium regions described in \cref{subsec:regions}. A schematic structure of the wave profile depending on the coordinate $\xi$ is given in \cref{fig.profile}. This provides a solution to our problem for the simplest case of a single oil component.

\section{General multi-component theory}
\label{sec.Ncomp}

We now show that a similar construction is possible in the general case of $N$ oil components. This includes a demonstration of the folding structure of the integral surface, and analysis in the equilibrium and out-of-equilibrium regions. Due to complexity of the system, only asymptotic analysis is possible, where we again assume that the parameters $\beta_i$, $\sigma$ and $\nu_{o_i}$ are small. The folding structure provides the resonance condition at the point separating the two regions, which brings a missing determining condition for the combustion wave. Note that this final condition is accurate, because it follows from the qualitative requirement that the wave profile passes through the fold singularity.

\subsection{Folding conditions}

Similarly to \cref{ExtraFS1}, with small parameters $\beta_i$, $\sigma$ and $\nu_{o_i}$, we approximate the fluxes \cref{NeqFOLD.psio} and gas speed \cref{eqCW.007BB} in the $N$-component formulation as
\begin{equation}
 \psi_o \approx
- \psi_g \left(\sum_{i=1}^N \frac{Y_i}{\nu_{o_i}}\right)
\left(\sum_{i=1}^N \frac{X_i}{\nu_{o_i}\beta_i}\right)^{-1},
\quad
 \psi_g =
 \frac{1}{\sigma}\left(1-\sum_{i=1}^N Y_i\right)^{-1},
 \quad
u\approx  \psi_g(1+\theta/\theta_0).
 \label{ExtraNC1}
\end{equation}
One can check that only leading-order terms are kept in these expressions. 
Since both $v/u$ and $\psi_o/u$ are small in \cref{eqCW.007B}, an approximation of $f_o$ at small values is justified. For example, let us take $f_o(\theta,\mathbf{X},s_o) = \frac{1}{2}f_2(\theta,\mathbf{X})s_o^2+o(s_o^2)$. 
Then Eqs.~\cref{eqCW.007B} and \cref{ExtraA4B}) for the fold singularity yield the approximate conditions
\begin{equation}
\textrm{fold:}\quad
s_o  \approx \frac{v}{f_2(\theta,\mathbf{X})u},
 \quad
\psi_o \approx -\frac{v^2}{2f_2(\theta,\mathbf{X})u}.
\label{ExtraNC3}
\end{equation}

Recall that Eq.~\cref{eqCW.007B} defines the integral hypersurface in $(2N+2)$-dimensional space $(\theta,\mathbf{X},\mathbf{Y},Y_k,s_o)$. The fold is defined as a singularity, when the integral surface is projected to the $(2N+1)$-dimensional space $(\theta,\mathbf{X},\mathbf{Y},Y_k)$ along the axis $s_o$. This fold is determined by the two equations~\cref{ExtraNC3} and represents a surface of dimension $2N$.

\subsection{Oil composition in equilibrium region}
\label{sec:compos}

We divide the wave profile into the equilibrium and out-of equilibrium regions.
On physical reasons, the out-of-equilibrium region is located upstream, $\xi < 0$.
Here the reaction and vaporization occur when the injected air contacts the oil.
Since the reaction and vaporization rates are large, this region is thin.
The wide equilibrium region follows downstream at $\xi > 0$, \cref{fig.profile}.

In the equilibrium region, conditions \cref{ExtraA7} imply the lack of oxygen and gas-liquid equilibrium, i.e.,
\begin{equation}
\textrm{equilibrium region:}\quad
Y_k = 0,\quad
Y_i = Y_i^{eq}(\theta,\mathbf{X}),\quad i = 1,\ldots,N,
\label{ExtraNC4}
\end{equation}
see \cref{eqWReq} and \cref{eqWReqB}. The condition of complete oxygen consumption, $Y_k = 0$, leads to the vanishing reaction rate $w_{r_i}$. The decrease of temperature along the profile in downstream direction implies that the condensation rate, $w_{v_i} < 0$, is small compared to its large out-of-equilibrium value, but it does not vanish. This consideration yields extra conservation laws that are valid only in the equilibrium region and can be derived as follows.

Taking $w_{r_i} = 0$ in \cref{NeqCW.Xi}, the combination of the resulting equations with \cref{NeqCW.Yi} yields
\begin{equation}
\frac{d}{d \xi}(X_i \psi_o+\beta_iY_i \psi_g) = 0,
 \quad i = 1,\ldots,N-1,
\label{ExtraNC5}
\end{equation}
which means that $N-1$ quantities in the parentheses do not change in the equilibrium region. 
Using \cref{ExtraNC4} and the downstream conditions \cref{eqCW.011}, \cref{eqCW.011fluxes} yields
\begin{equation}
X_i \psi_o + \beta_i \psi_g Y_i^{eq}(\theta,\mathbf{X})
= X_i^{ini}\psi_o^d  + \beta_i \psi_g^d Y_i^{ini},
 \quad i = 1,\ldots,N-1.
 \label{eqER.int}
\end{equation}
The downstream-state fluxes $\psi_o^d$ and $\psi_g^d$ are given explicitly in \cref{eq.psiodA}, while the fluxes $\psi_o$ and $\psi_g$ are expressed as functions of $\theta$ and $\mathbf{X}$ from \cref{NeqFOLD.psio} and \cref{ExtraNC4}. 
As a result, $N-1$ equations in \cref{eqER.int} depend on $N$ variables $(\theta,\mathbf{X})$.
Solving these equations (e.g., numerically) provides the dependence of oil composition on temperature: $\mathbf{X} = \mathbf{X}(\theta)$.

\subsection{Temperature profile}
\label{sec:tp}

As soon as we have all variables expressed in terms of temperature in the equilibrium region, the temperature $\theta$ as a function of the coordinate $\xi$ can be obtained from Eq.~\cref{ExtraA1}. Using \cref{ExtraNC4}, we write this equation as
\begin{equation}
 \frac{d\theta}{d\xi} =
 (-v + \alpha_o\psi_o + \alpha_g\psi_g)\theta.
 \label{eqONE.tExt}
\end{equation}
The fluxes are expressed from \cref{NeqFOLD.psio} with conditions \cref{ExtraNC4} as
\begin{equation}
 \psi_o =
 - \left(\frac{Y_k^{inj}}{\sigma}
  +\psi_g
 \sum_{i=1}^N \frac{Y_i^{eq}}{\nu_{o_i}}
\right)
 \left(\sum_{i=1}^N \frac{X_i}{\nu_{o_i}\beta_i}\right)^{-1},
  \quad
 \psi_g =
 \frac{1}{\sigma}\left(1-\sum_{i=1}^N Y_i^{eq}\right)^{-1},
 \label{ExtA1}
\end{equation}
where $Y_i^{eq}(\theta,\mathbf{X})$ are given functions and the oil composition $\mathbf{X}(\theta)$ is determined as explained in \cref{sec:compos}. This defines a scalar ordinary differential equation in \cref{eqONE.tExt}, which can be integrated numerically with the initial condition $\theta = \theta^u$ at $\xi = 0$.

We do not have a simple argument that justifies that the right-hand side in \cref{eqONE.tExt} is everywhere negative, as we did in the case of a single component. However, numerical simulations in \cref{sec:numer} indeed provide a monotonously decreasing profile of the temperature with $\xi$.

\subsection{Resonance point}

The wave profile must follow the vector field induced by balance laws on the folded integration surface in the space $(\theta,\mathbf{X},\mathbf{Y},Y_k,s_o) \in \mathbb{R}^{2N+2}$, see \cref{subsec:overall}. 
The fold on this surface is defined as a singularity under projection to the space $(\theta,\mathbf{X},\mathbf{Y},Y_k)$. 
As we explained in \cref{subsec:asym}, the profile can be split into the equilibrium and out-of-equilibrium parts. 
In the equilibrium region, all variables were expressed (for given wave speed $v$) as functions of temperature in \cref{sec:compos}. 
In the out-of-equilibrium region, the temperature is approximately constant, see \cref{ExtraA8}, and the profile in the space $(\theta,\mathbf{X},\mathbf{Y},Y_k)\in \mathbb{R}^{2N+1}$ is given by equations \cref{NeqCW.Xi}--\cref{NeqCW.Yk}.

Due to immense complexity of the multi-component problem, it is hard to study the folded integration surface with the vector field for the wave profile in the general case. It is however possible to formulate some generic transversality assumption that can be checked a posteriori. As such conditions, we propose to assume the transversality of the equilibrium and out-of-equilibrium parts of the wave profile to the folding surface in the projection to $(\theta,\mathbf{X},\mathbf{Y},Y_k)$. As there is no direct functional relation among these three objects (folding line and two profile sections), this transversality condition is natural.

The transversality assumption for each region (equilibrium and out-of-equilibrium) implies that the corresponding section of the wave profile cannot be extended beyond the folding line, see \cref{fig.fold}. This, in turn, means that the point, which connects the two regions, is a fold singularity. By continuity, the fold condition \cref{ExtraA4B} is satisfied at this point together with the conditions \cref{ExtraNC4} and \cref{ExtraA8}. For the latter, the temperature value corresponds to the upstream (hot) state, $\theta \approx \theta^u$. In summary, we have the resonance point at
\begin{equation}
 \theta = \theta^u,\quad
 \mathbf{X} = \mathbf{X}(\theta^u),\quad
 Y_i = Y_i^{eq}(\theta^u,\mathbf{X}(\theta^u)),\quad Y_k = 0,
 \label{eqAP.v}
\end{equation}
where $\mathbf{X}(\theta)$ is determined by Eq.~\cref{eqER.int} as explained in \cref{sec:compos}.

The two equations valid at the resonance point are \cref{eqCW.007B} and \cref{ExtraA4B}, where $\psi_o$, $\psi_g$ and $u$ must be substituted from \cref{NeqFOLD.psio} and \cref{eqCW.007BB}. With \cref{eqAP.v} and $\theta^u$ expressed from \cref{eq.tuF}, these two equations contain only two unknowns: the wave speed $v$ and the oil saturation at the fold $s_o = s_o^f$. Thus, the folding singularity provides the full system of determining equations for the wave speed $v$. Having found the wave speed $v$, the other wave parameters are obtained as explained in \cref{sec:Conl}. Namely, three equations in \cref{eq.tuF} and \cref{eq.psiodA} with the fluxes expressed from \cref{eqCW.011fluxes} can be solved with respect to the three unknowns: $\theta^u$, $s_o^d$ and $u_g^d$.

We see that the complete system of equations follows from the transversality assumption at the folding singularity, providing the recipe for solving our problem in the general $N$-component case. We cannot guarantee that this solution exists for any problem parameters and model functions, such as $f_o(\theta,\mathbf{X},s_o)$, $Y_i^{eq}(\theta,\mathbf{X})$, etc. However, the example shown below indicates that this solution may be valid for a large family of multi-component models.

\section{Numerical example}
\label{sec:numer}

For a specific example, we need to specify the functions $f_o$ and $Y_i^{eq}$, as well as all problem parameters.
Following~\cite{kokubun2016}, based on the Clausius-Clapeyron relation with Raoult's law, we define
\begin{equation}
Y_i^{eq}
=
{X}_i
\frac{P_{atm}}{P}
\exp
\left(
-\frac{Q_{v_i}}{R}\left(\frac{1}{T}
-\frac{1}{T_{n_i}}\right)
\right), \ \ \ i=1,\ldots,N,
\label{eq2.CC}
\end{equation}
where $T_{n_i}$ is the (normal) boiling point of the pure $i$th component measured at atmospheric pressure  $P_{atm}$.
The dimensionless expression is given by
\begin{equation}
Y_i^{eq}(\theta,X_i)
=
{X}_i \mathcal{Y}_i(\theta),
\label{eqND.Yheqb}
\end{equation}
where we defined the functions $\mathcal{Y}_i(\theta)$ as
\begin{equation}
 \mathcal{Y}_i(\theta)
 =
 \mbox{exp}\left(\frac{\theta_{h_i}}{\theta_0 + 1 + \Delta \theta_i} - \frac{\theta_{h_i}}{\theta_0 + \theta}\right),
 \label{eqND.c}
\end{equation}
and
\begin{equation}
\theta_{h_i} = \frac{Q_{v_i}}{R\Delta T^*}, \ \ \
\Delta\theta_i = \frac{T_{b_i} - T_{b_1}}{T_{b_1} - T_{ini}}.
\label{eqND.thetah}
\end{equation}
We chose the characteristic temperature $T^*$ as the boiling temperature of the first component, i.e., $T^* = T_{b_1}$.
For the relative permeability function, we take (see e.g.~\cite{kokubun2016})
\begin{equation}
f_o(s_o,T,{X}_i) = \frac{k_o/\mu_o}{k_o/\mu_o + k_g/\mu_g},\quad
k_o = \frac{(s_o - s_{or})^2}{(1 - s_{or})^2},\quad
 k_g = s_g^2,
\label{eqM.12}
\end{equation}
where $k_o = 0$ for $s_o \le s_{or}$ with the residual oil saturation $s_{or}$, and the oleic and gas viscosities are
\begin{equation}
\mu_o = \left(\sum_{i=1}^N\frac{X_i}{\mu_i^{1/4}}\right)^{-4},
\quad
\mu_i = \mu_{i_0}~\mbox{exp}\left(-\frac{E_i}{R T}\right),
\quad
 \mu_g = \frac{7.5}{T + 120}\left(\frac{T}{291}\right)^{3/2}.
\label{eq.VIS}
\end{equation}
In Eqs. \cref{eq.VIS}, $E_i$ is the activation energy for the viscosity of the $i$th component in the
oleic phase and $\mu_{i_0}$ is a reference viscosity.
An explicit form of the reaction and vaporization terms is not necessary for our analysis,
but one can see~\cite{gargar2014jpm} for specific examples.

In order to present some quantitative results of our model, we consider the case $N = 3$ oil components.
Specifically, we consider a mixture of components representative for heptane ($X_1$), decane ($X_2$) and hexadecane ($X_3$).
The parameters that describe the light oil reservoir are given in \cref{table1}.
We assume a reservoir pressure of $P = 80$~atm, such that the boiling temperatures are $T_{b_1} = 644$~K, $T_{b_2} = 913$~K and $T_{b_3} = 1554$~K.
These boiling temperatures yield $\Delta\theta_1 = 0$, $\Delta\theta_2 = 0.832$ and $\Delta\theta_3 = 2.811$.
The dimensionless parameters are given in \cref{table2}.

First, lets consider a homogeneous initial oleic mixture, i.e., $X_1^{ini}=X_2^{ini}=X_3^{ini}=1/3$.
We solve Eqs. \cref{eqER.int} in order to obtain the compositions $X_i$ in terms of the temperature $\theta$.
Then, we solve Eqs. \cref{ExtraFS8} evaluated at the resonance point, i.e., at $s_o^r$ and $\theta^u$, and where the gas velocity $u$ and the
oleic flux $\psi_o$ are given by Eqs. \cref{ExtraNC1}.
From the solutions of Eqs. \cref{ExtraFS8} we obtain the oil saturation at the resonance and the upstream temperature, which determines the wave speed $v$
from Eq. \cref{eq.tuF}.
With the wave speed we can evaluate the first Eq. \cref{ExtraFS8} at downstream, $\theta=0$, where $\psi_o^d$ is given by Eq. \cref{eq.psiodA},
in order to find the downstream saturation $s_o^d$.
Then, all unknowns of the limiting states are obtained, as well as the wave speed.

\begin{table}[h!]
\caption{Parameters used in numerical examples~\cite{kokubun2016}.}
\label{table1}
\begin{center}
\begin{tabular}{lll|lll|lll}
$c_g$    & = & $29$ J$/$ mol K        & $R$              & = & $8.314$ J$/$K$/$mol      & $\mu_{o_3}$     & = & $2.33\times10^{-2}$ Pa s\\
$C_m$    & = & $2.86$ MJ$/$ m$^3$ K   & $s_o^r$          & = & $0.1$                    & $\nu_{o_1}$     & = & $0.09$\\
$C_o$    & = & $1.53$ MJ$/$ m$^3$ K   & $T_{ini}$        & = & $320$ K                  & $\nu_{o_2}$     & = & $0.03$\\
$E_1$    & = & $8364$ J$/$mol         & $u_{inj}$        & = & $8\times10^{-4}$ m$/$s   & $\nu_{o_3}$     & = & $0.065$\\
$E_2$    & = & $10185$ J$/$mol        & $Y_k^{inj}$      & = & $0.21$                   & $\rho_{o_1}$    & = & $6826$ mol$/$m$^3$\\
$E_3$    & = & $17742$ J$/$mol        & $\lambda$        & = & $3$ W/m K                & $\rho_{o_2}$    & = & $5130$ mol$/$m$^3$\\
$P$      & = & $8\times10^6$ bar      & $\mu_{o_1}$      & = & $1.32\times10^{-2}$ Pa s & $\rho_{o_3}$    & = & $5330$ mol$/$m$^3$\\
$Q_{r_i}$& = & $4\times10^5$ J        & $\mu_{o_2}$      & = & $1.42\times10^{-2}$ Pa s& $\varphi$       & = & $0.3$\\
$Q_{v_i}$& = & $31.8$ kJ$/$mol        &                  &&                             &                 &&\\
\end{tabular}
\end{center}
\end{table}

\begin{table}[h!]
\caption{Dimensionless parameters.}
\label{table2}
\begin{center}
\begin{tabular}{lll|lll}
$\alpha_g$     & = & $0.0131$        & $\theta_{h_i}$ & = & $11.8112$\\
$\beta_1$      & = & $0.4405$        & $\theta_0$     & = & $0.9882$ \\
$\beta_2$      & = & $0.5862$        & $\sigma$       & = & $0.117$\\
$\beta_3$      & = & $0.5642$        & &&\\
\end{tabular}
\end{center}
\end{table}

\begin{figure}
 \begin{center}
  \includegraphics[width=0.45\linewidth]{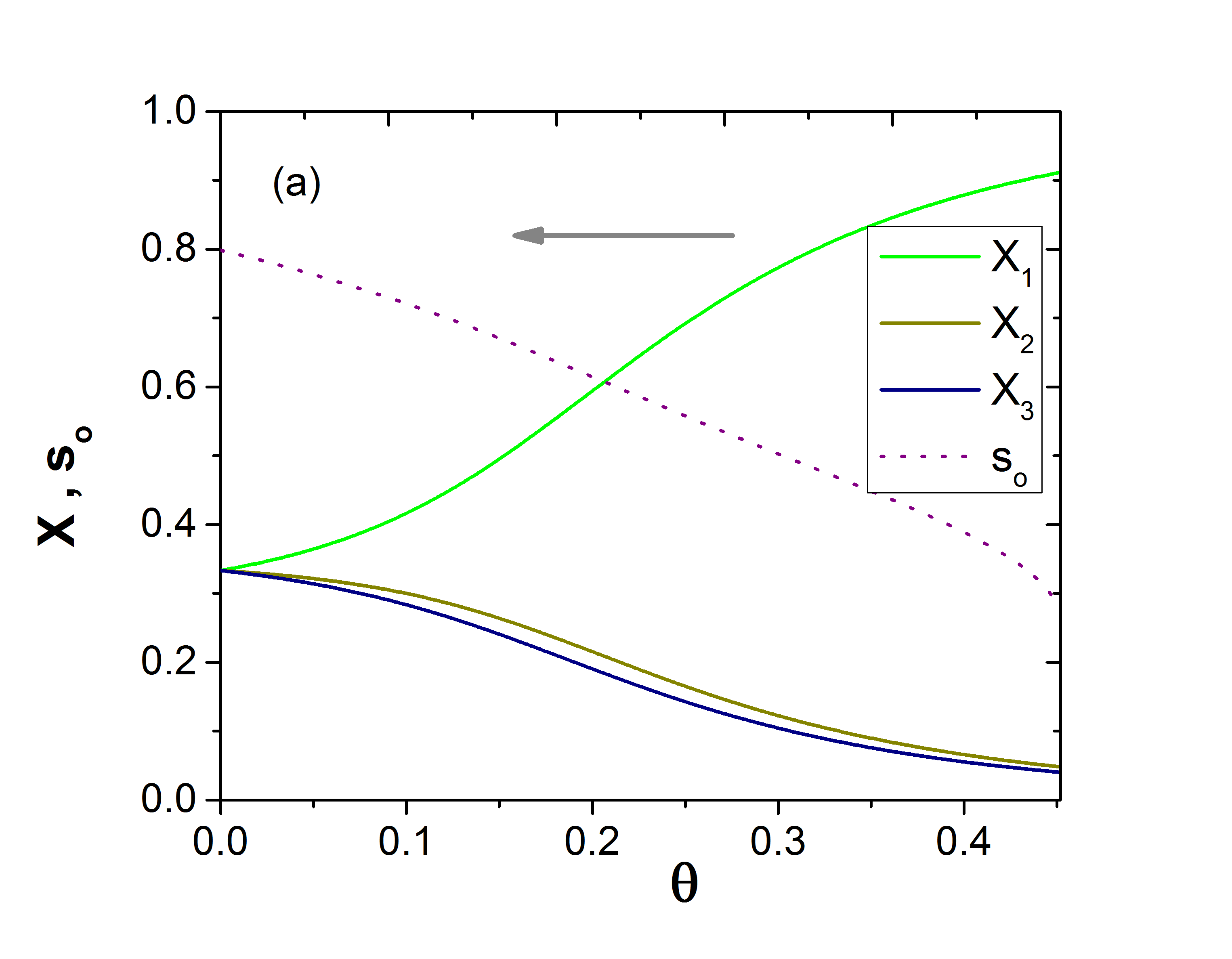}
  \includegraphics[width=0.45\linewidth]{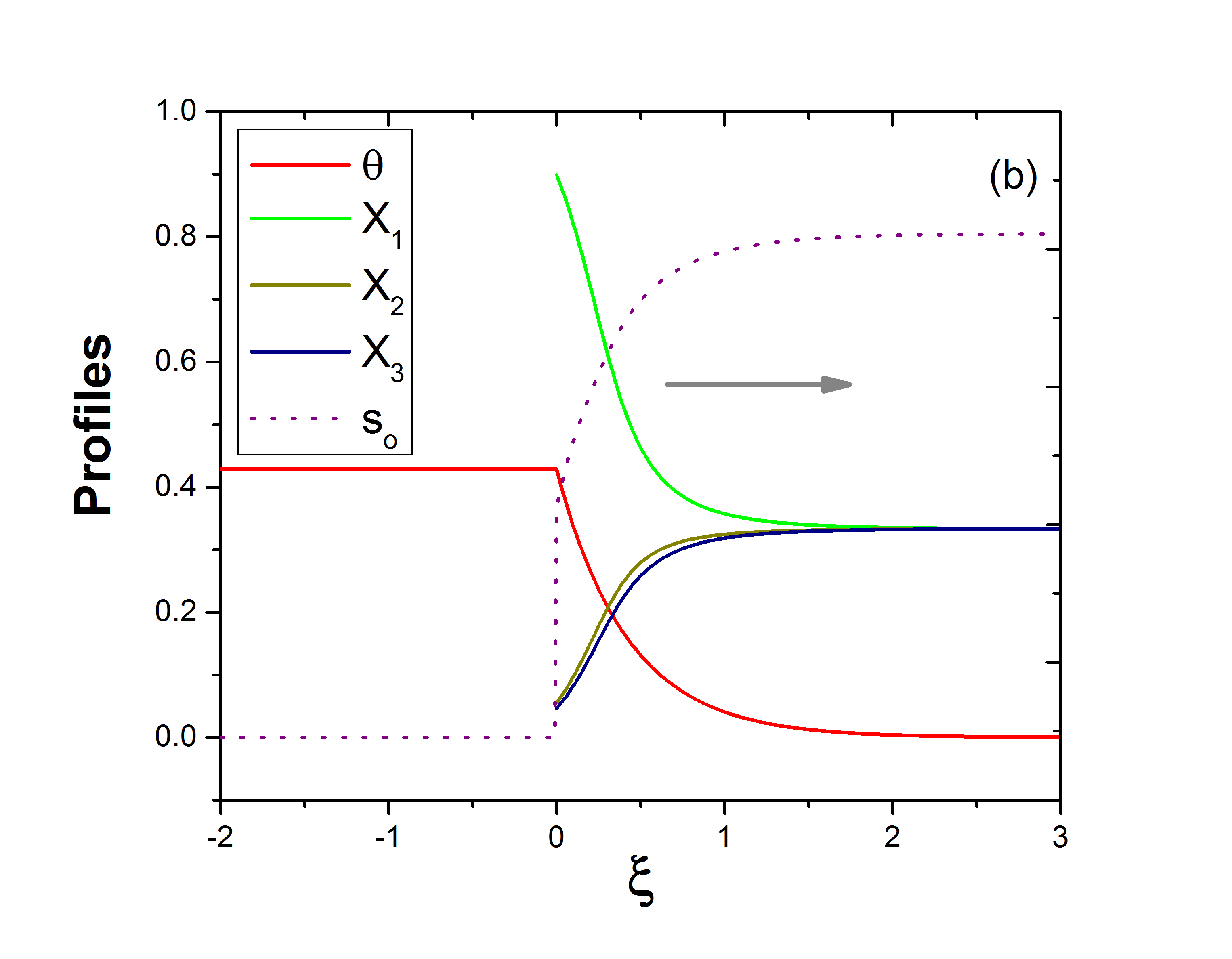}
  \end{center}
  \vspace{-5mm}
  \caption{Profile of the wave in terms of (a) $\theta$ and (b) $\xi$. The arrow indicates the downstream direction. The composition
is not defined for $\xi<0$ in (b), where $s_o = 0$.}
 \label{fig.xi}
\end{figure}

\begin{figure}
 \begin{center}
  \includegraphics[width=0.4\linewidth]{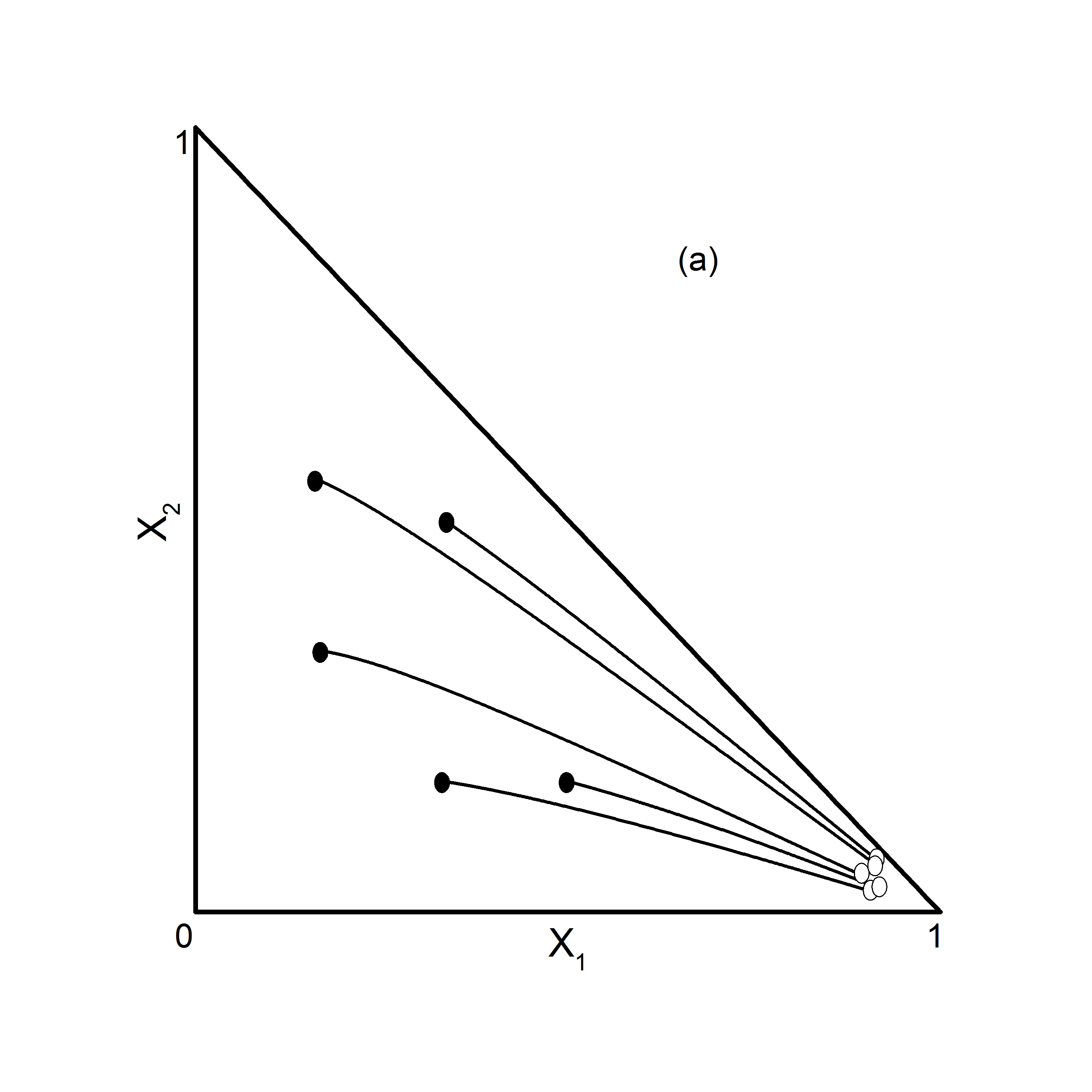}
  \includegraphics[width=0.4\linewidth]{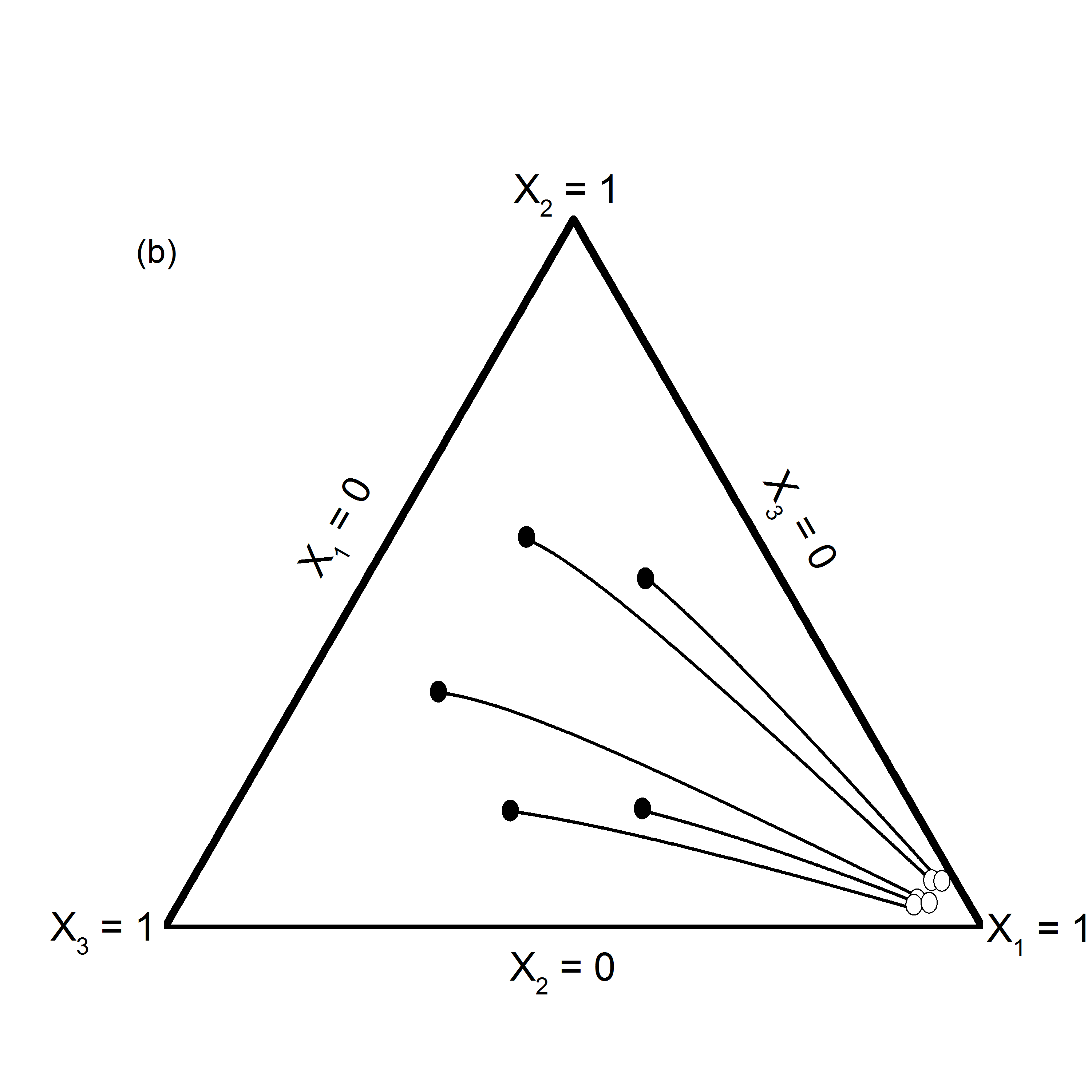}
  \end{center}
  \vspace{-10mm}
  \caption{Plots for the oil composition along the wave profile for five different initial compositions in (a) orthogonal coordinates and (b) ternary plot.
Open dots refer to the resonance point, while closed dots to the initial composition on the downstream side.}
 \label{fig.triangle}
\end{figure}

In \cref{fig.xi}(a) we show the composition $X_1,X_2,X_3$ and the oil saturation $s_o$ in the wave
with respect to the dimensionless temperature $\theta$, while in \cref{fig.xi}(b) we show the profiles with respect to $\xi$, indicating
the non-equilibrium, $\xi<0$, and equilibrium, $\xi>0$ regions.
The arrows indicate the direction of the wave propagation.
The temperature profile $\theta(\xi)$ along the wave shown in \cref{fig.xi}(b) is obtained by integrating Eq. \cref{eqONE.tExt} and
it confirms that the temperature profile decays monotonously in the downstream direction.
From \cref{fig.xi}(a) and \cref{fig.xi}(b) one can see that the heavier component is expelled from the wave in the downstream direction.	
This counterintuitive property results in leaving the more volatile component in the wave, and particularly at the resonance point.
Hence, the important physical processes occurring in the resonance point, i.e., oxidation and vaporization, are largely determined by the
lightest component of the oleic mixture.
In our example, the properties of heptane (lighter component) determines the wave parameters, i.e., combustion temperature, wave speed and downstream
saturation.
The same result can be seen in Figs. \ref{fig.triangle}, where we show the change in composition along the wave, from the resonance point (open
dots) to the downstream side (closed dots) for five different initial compositions.
In the left panel (a), we show the orbits along the wave in orthogonal coordinates $X_1$ and $X_2$, while in the right panel (b) we show the corresponding ternary plot.
The representation in a ternary plot is usual in multiphase flows in porous media, see, for instance \cite{marchesin2001wave,azevedo2014uniqueness}.
One can see that at the resonance point the oil is mostly composed by the lightest component $X_1$, for all initial oleic compositions.

The results shown in \cref{fig.xi} and \cref{fig.triangle} show that the composition at the resonance point is dominated by the lightest component
of the oil, regardless of the initial oleic composition.
Therefore, if the lightest component is kept fixed, it is expected that changing the heavier components will not change significantly the results.
In fact, changing $\Delta\theta_3$ from $2.811$ to $26.664$ yields a difference in the wave speed of less than $0.1\%$.

\section{Conclusions}
\label{sec.conc}

A profile of a nonlinear traveling wave is associated with a heteroclinic orbit for a respective vector field.
For a  system of balance laws in one space dimension, such a vector field is defined implicitly and, therefore, may have singularities (folds) at points,
where equations cannot be resolved with respect to derivatives.
Since a generic orbit cannot pass the folding line, extra conditions are imposed on the traveling wave at the singular point.
Examples of such waves though relatively rare are known for long time, e.g.,
the pathological detonation~\cite{fickett2011detonation,sharpe1999structure,kulikovskii2005propagation}.
The mathematical theory is developed for a single balance law~\cite{harterich2003viscous}.
However, this theory does not generalize directly to systems of balance laws and, apart from the studies for specific models,
the full mathematical theory is still missing, see, e.g.,~\cite{harterichtravelling,sotomayor2001impasse}.

In this paper, we study one more example of this singularity in a combustion wave.
This wave develops in two-phase flow in porous medium modeling light oil extraction by air injection.
A general $N$-component oil model is considered, represented by $2N+3$ balance laws coupled through filtration,
reaction and vaporization/condensation terms.
As the main result, this work shows that the singularity represents a clue to the analytical solution for the combustion wave,
despite a tremendous complexity of the model.
The mathematical technique that we propose is based on splitting the wave profile into the equilibrium and out-of-equilibrium regions,
and identifying the fold singularity in between.
Such splitting is approximate: it is valid asymptotically in the limit of large source terms.
With the singularity analysis, we provide a full set of determining equations for all wave parameters.

For applications, our work provides a combustion wave solution suitable for a large class of models.
It also reveals some important properties of the solution: the oil composition becomes much lighter in the region of high temperatures,
with a weak dependence on initial concentrations.
The key role played by the fold singularity in the solution to our problem justifies further development of the mathematical theory
focused on singular wave profiles for general systems of balance laws.

\bibliographystyle{siamplain}
\bibliography{refs}
\end{document}